\documentclass{aa}  

\makeatletter
\renewcommand*\aa@pageof{, page \thepage{} of \pageref*{LastPage}}
\makeatother

\usepackage{graphicx}
\usepackage{rotating}
\usepackage{txfonts}
\usepackage{color}
\usepackage{natbib}
\usepackage{amsmath}
\usepackage{hyperref}
\usepackage{siunitx}
\usepackage{tablefootnote}
\bibpunct{(}{)}{;}{a}{}{,} 
\newcommand{\diamonds}{\textsc{Diamonds}}           
\newcommand{\numax}{\nu_\mathrm{max}}

\newcommand{\logg}{\log\text{g}}
\newcommand{\dnu}{\Delta\nu}
\newcommand{\kepler}{{{\it Kepler}}}
\newcommand{\apollo}{\texttt{APOLLO}}
\DeclareSIUnit\parsec{pc}  
\DeclareSIUnit\year{yr}  
\usepackage{float}
\usepackage{natbib}
\usepackage{txfonts}
\begin{document}

   \title{Searching for solar-like oscillations in pre-main sequence stars using \apollo \thanks{Partly based on observations made with ESO Telescopes at the La Silla
Paranal Observatory under programme 097.C-0040(A).}}

  \subtitle{Can we find the young Sun?}

   \author{M. M\"ullner\inst{1} \and 
            K. Zwintz\inst{1} \and
          E. Corsaro\inst{2} \and
          T. Steindl\inst{1} \and
          I. Potravnov\inst{3} \and
          E. W. Guenther\inst{4} \and 
          A. Kniazev\inst{5,6,7} \and
          V. Gvaramadze\inst{7,8}
                  }

   \institute{Institut f\"ur Astro- und Teilchenphysik, Universit\"at Innsbruck, Technikerstra{\ss}e 25, A-6020 Innsbruck\\
              \email{marco.muellner@student.uibk.ac.at} \and
    INAF - Osservatorio Astrofisico di Catania, via S. Sofia 78, 95123, Catania, Italy
                \and
    Institute of Solar-Terrestrial Physics, Siberian branch of Russian Academy of Sciences, Lermontov Str. 126A, 664033, Irkutsk, Russia
    \and
    Th\"uringer Landessternwarte Tautenburg, Sternwarte 5, 07778 Tautenburg, Germany
    \and 
    South African Astronomical Observatory, PO Box 9, 7935 Observatory, Cape Town,South Africa
    \and
    Southern African Large Telescope Foundation, PO Box 9, 7935 Observatory, Cape Town, South Africa
    \and
    Sternberg Astronomical Institute, Lomonosov Moscow State University, Moscow 119234, Russia
    \and
    Space Research Institute, Russian Academy of Sciences, Profsoyuznaya 84/32, 117997 Moscow, Russia}

   \date{}

   \date{Received; accepted}

 
  \abstract
   {In recent years, our understanding of solar-like oscillations from main sequence to red giant stars has improved dramatically thanks to pristine data collected from space telescopes. One of the remaining open questions focuses around the observational identification of solar-like oscillations in pre-main sequence stars.
   }
   {We aim to develop an improved method to search for solar-like oscillations in pre-main sequence stars and apply it to data collected by the Kepler K2 mission.}
   {
    Our software {\texttt{APOLLO}} includes a novel way to detect low signal-to-noise ratio solar like oscillations in the presence of a high background level.
  }
   {By calibrating our method using known solar-like oscillators from the main \kepler\ mission, we apply it to T Tauri stars observed by Kepler K2 and identify several candidate pre-main sequence solar-like oscillators. }
   {We find that our method is robust even when applied to time-series of observational lengths as short as those obtained with the TESS satellite in one sector. We identify EPIC\,205375290 as  a possible candidate for solar-like oscillations in a pre-main sequence star with $\numax \simeq 242\,\mu$Hz. We also derive EPIC\,205375290's fundamental parameters to be $T_\mathrm{eff}$ = 3670$\pm$180\,K, log\,$g$ = 3.85$\pm$0.3,  $v$sin$i$ = 8 $\pm$ 1 km\,s$^{-1}$, and about solar metallicity from a high-resolution spectrum obtained from the Keck archive.}

   \keywords{Asteroseismology -- Methods: data analysis -- Stars: interiors -- Stars: pre-main sequence -- Stars: solar-type -- Stars: individual: EPIC\,205375290
               }

   \maketitle

%

\section{Introduction}

Asteroseismology is the most powerful method to describe the inner structure of stars. 
In recent years, several space missions have delivered pristine photometric data that significantly advanced our knowledge about stellar interiors and the physical processes acting inside stars.

Two of the pioneering space telescopes for asteroseismology, were the Canadian microsatellite MOST \citep[Microvariability and Oscillations of Stars;][]{walker03} and the french-led mission CoRoT \citep[Convection, Rotation, and Planetary Transits;][]{auvergne09}. They were followed by the NASA \kepler\,\citep{Borucki2010,Koch2010} space telescope. After the failure of two reaction wheels in 2013, the mission continued as K2 \citep{Howell2014} with a different observing strategy.

 Solar like oscillations are caused by turbulent convective motions near the star's surface and show p-modes in its frequency spectrum \citep{Chaplin2013,Garcia2019}. These types of oscillations have very small amplitudes, making space missions the primary driver for detecting them. The previously mentioned space missions, starting with CoRoT \citep[e.g.,][]{michel2008,deridder2009}, were especially impactful in the search for solar like oscillators.

The largest progress by far was made by studying solar-like oscillations in main sequence (MS) and red giant stars \citep[e.g.][]{Lund2017,Bastien2013,Huber2011,antoci2011}. These p-mode pulsations are driven by the outer convective layer, have pressure as their restoring force, and are described by two typical parameters: the large frequency separation ($\Delta\nu$) and the frequency of maximum power ($\numax$).
Scaling relations connect these two asteroseismic observables with the stellar fundamental parameters \citep{KjeldsenBedding1995,Huber2011} and are based on the observed values for the Sun.

All of the present observational studies of solar-like oscillations describe stars starting from the main sequence phase of the stellar evolution.
For a complete evolutionary picture, the connection of the early phases of stellar evolution with the later phases is essential. We aim to understand the properties of "young Suns'' in terms of activity and rotation \citep[e.g.][]{Frohlich2012} to be able to trace the Sun back to its initial conditions. In this, the next logical step is to investigate the oscillatory past of the Sun and solar-like stars by searching for solar-like oscillations in pre-main sequence (pre-MS) stars.

The presence of solar-like oscillations in pre-MS stars was proposed by \cite{Samadi2005} and \cite{Pinheiro2008} based only on theoretical calculations. Until today no pre-MS solar-like oscillators were detected observationally. This is due to the high degree of activity present in young stellar objects, which introduces a high background and obscures any oscillation signal, but also due to the suppression of oscillation mode amplitudes due to the presence of magnetic activity \citep{Chaplin2011,Bonanno2014}, that further reduces the signal to noise ratio, making detections very difficult. Consequently, one important aspect for our search for pre-MS solar like oscillators is the formulation of the background. Some recent achievements in the background formulation have been discussed by e.g. \cite{Mathur2011a,Karoff2012,Kallinger2014,Kallinger2016,Corsaro2017}.

Another reason why no pre-MS solar like pulsators have been discovered observationally so far, is the lack of 
high-precision photometric time-series of young stellar objects preferably from space and with sufficiently long time bases. As the main \kepler\, mission was pointing away from young star forming regions on purpose, the current maximum observing lengths for pre-MS pulsators is on the order of about 80 days from Kepler K2 \citep{Howell2014} and up to about 105 days as obtained using data collected by the NASA mission TESS \citep{Ricker2014}. Also, these relatively short observational time bases have an impact on the noise sources that therefore have to be treated properly when searching for low-amplitude stochastic oscillations. 

T Tauri stars are the main candidate pre-MS objects to search for solar-like oscillations. These young stars, which are usually of spectral types late F, G, K, and M, have recently become visible in the optical and have masses $\lesssim 2\text{M}_\odot$. Their light variability can vary over a large range of time scales, from minutes to decades, and show both regular and irregular behaviour \citep[e.g.,][]{alencar2010,cody2014}. 
The size and location of their outer convective regions make these objects resemble solar-like stars. 

Space missions such as \kepler, TESS, and the future ESA mission PLATO \citep[][{expected launch in 2026}]{Rauer2014} have already brought asteroseismology into the era of big data: 
With sample sizes on the order of a minimum of half a million objects, new methods and pipelines are required for the analysis of stellar oscillations that allow for automated processing. These methods both need to be computationally efficient, to process such large datasets in a reasonable amount of time, and  have to yield reliable results, possible through rigorous statistical methods such as the Bayesian approach used in the present work. Previous works from \cite{Campante2016} and \cite{Schofield2019} have already discussed the detectability of solar-like oscillations from TESS observations, both from a 
theoretical standpoint and from the nominal photometric performance of TESS.

In this work, we present the \apollo\, (Automated Pipeline for sOlar-Like Oscillators, available online\footnote{https://github.com/MarcoMuellner/APOLLO}) pipeline: additionally to computing the background, it includes a novel way to detect low signal-to-noise ratio solar-like oscillators by applying model comparisons in a ``blind'' way. For this, a Bayesian approach based on \diamonds\ \citep{corsaro2014} is used, which allows to quantify the significance of a signal in a given dataset. We apply  \apollo\, to known solar-like oscillators and calibrate the method. In a final step, we investigate data from the K2 mission for potential pre-MS solar like oscillators with \apollo\, and present the first candidates.


\section{Target selection}
\subsection{Main sequence and post-main sequence star sample}
For the development and calibration of our method, we selected 1119 solar-like oscillators observed by the \kepler\, satellite, provided through the KASC\footnote{http://kasoc.phys.au.dk/} collaboration \citep{Kjeldsen2010}. Of these 1119 objects, 1071 are obtained in long cadence \citep[LC; sampling time $\Delta t= 29.45$\,min][]{Jenkins2010}, representing a sample of evolved red-giant stars with available fundamental properties from photometry, as well as spectroscopic parameters from the APOKASC catalogue \citep{Pinsonneault2014}; 48 targets are short cadence \citep[SC; sampling time $\Delta t=59.89$\,s][]{Gilliland2010}, giving us our subset of main sequence and subgiant stars from the LEGACY sample \citep{Lund2017}, for which fundamental asteroseismic parameters are available from the literature.

All the red giants in our sample have been observed for at least two years, with $60$\% having four years of nearly continuous measurements. We did not explicitly impose limiting values of effective temperature and magnitude on any of the stars during our selection, apart from the constraints of the the APOKASC catalog. Accordingly, the frequency spectra of the selected red giants cover $\nu_\text{max}$ values ranging from $1\,\mu$Hz to $246\,\mu$Hz.

The SC sample consists of both ``Simple'' and ``F-Type'' stars, following \cite{Appourchaux2012}. These stars have an observation baseline of at least $370$\,d, with $59$\% showing a baseline of $1095$\,d ($\sim 3$\,yr). The frequency ranges for these stars cover $\nu_\text{max}$ values between $885\,\mu$Hz and $3456\,\mu$Hz, representing our main sequence and subgiant sample. Additionally, effective temperatures and \kepler\, magnitudes, $Kp$, for our full sample of 1119 stars were extracted from the \kepler\, input catalogue \citep{Brown2011} to estimate $\numax$.

\subsection{Candidate pre-MS solar-like stars}
For our search for pre-MS solar like oscillators, we identified 135 candidate pre-MS solar like oscillators 
in the data from the K2 mission Campaign 2, which partially observed the Upper Scorpius (USco) subgroup. This group is one of three that belong to the Scorpius-Centaurus OB Association (Sco-Cen) and also the youngest one with 10\,$\pm$\,3\,Myr.
The 135 stars selected all show an effective temperature between 4500 and 7000\,K, are members of USco and, consequently, are identified as pre-MS stars. To obtain the surface gravity, $\log \text{g}$, effective temperature, $T_\text{eff}$, and radii of these stars, used for the identification process of solar-like oscillation, we use the Ecliptic Plane Input Catalog (EPIC) by \cite{Huber2016}.

\section{Methodology}
The methodology used in the \apollo\, pipeline aims to determine the asteroseismic parameters $\nu_\text{max}$ and $\dnu$ of solar-like oscillators by exploiting a Bayesian approach. 
For this, we only consider the light curve, effective temperature and magnitude of a given star. The following subsections describe our method in detail.

\subsection{Data preparation}\label{reduction}
The light curves used in this work were provided by KASC and directly downloaded from their website, which already includes the reduction using the KASOC filter \citep{Handberg2014}.

As our first step, we create a histogram of the relative flux values which is then fitted with a normal distribution, providing its mean and standard deviation, $\sigma$. All data points outside $4 \sigma$ are then removed from the light curve and gaps longer than three days are linearly interpolated connecting the two points adjacent to the gap.

For these resulting light curves we compute the power spectral density (PSD) using the Lomb-Scargle periodogramm \citep{Lomb1976,Scargle1982}, normalized using Parseval's theorem and finally converted to the PSD by dividing through the integral of the spectral window. This PSD is the main input for our further analysis.

\subsection{Bayes theorem and model comparison}\label{sec:Bayes_theorem_model_comp}

\begin{table}
\caption{Empirical scale for comparison of the strength of evidence to determine the Bayes factor (\citealt{Trotta2006}). $O_{ij}$ is the odds ratio.}       
\label{Tab:Evidence}      
\centering                                      
\begin{tabular}{c c c c}          
\hline\hline                        
\textbar{}$\ln{ O_{ij}}$ \textbar{}& Odds    & Probability & Strength of evidence  \\ 
\hline                                   
<1.0                       & <3 : 1   & <0.750       & Inconclusive          \\
1.0                       & 3 : 1   & 0.750       & Weak evidence         \\
2.5                       & 12 : 1  & 0.923       & Moderate evidence     \\
5.0                       & 150 : 1 & 0.993       & Strong evidence       \\
\hline                                             
\end{tabular}
\end{table}

The core of our analysis is based on the Bayes theorem, which is formulated as

\begin{equation}\label{bayes_theorem}
    p ( \boldsymbol { \theta } | d , \mathcal { M } ) = \frac { \mathcal { L } ( \boldsymbol { \theta } | d , \mathcal { M } ) \pi ( \boldsymbol { \theta } | \mathcal { M } ) } { p ( d | \mathcal { M } ) } \, .
\end{equation}

where $\boldsymbol { \theta }$ is the parameter vector consisting of the free parameters of the model $\mathcal { M }$ applied to the data set $d$, which is in our case the stellar PSD. The likelihood function $\mathcal { L } ( \boldsymbol {\theta} | d , \mathcal { M })$ represents the way the data is sampled. To describe Fourier power spectra, (i.e. our PSD), it is necessary to adopt the \textit{exponential likelihood} as introduced by \citet{Duvall1986,Anderson1990}. The parameter $\pi ( \boldsymbol { \theta } | \mathcal { M })$ denotes the prior probability density function (PDF), defined through their respective prior distributions. The denominator in Eq.~(\ref{bayes_theorem} is the Bayesian evidence (also called marginal likelihood or model performance), giving us a quantitative evaluation of the quality of the model in light of the data. It is defined by

\begin{equation}\label{eq:evidence_equation}
  E  \equiv p ( d | \mathcal { M } ) = \int _ { \Sigma _ { \mathcal { M } } } \mathcal { L } ( \boldsymbol { \theta } ) \pi ( \boldsymbol { \theta } | \mathcal { M } ) \mathrm { d } \boldsymbol { \theta }
\end{equation}

This value is of critical importance for the model comparison. Assuming two distinct models $\mathcal { M }_i$ and $\mathcal { M }_j$, we can apply these two models to our data set d. We can then take their respective evidence $E_i$ and $E_j$, and compute the odds ratio $O _ { i j }$, defined by

\begin{equation}\label{eq:bayes_factor}
O _ { i j } \equiv \frac { p \left( \mathcal { M } _ { i } | d \right) } { p \left( \mathcal { M } _ { j } | d \right) } = \frac { E_{ i } } {E_{ j } } \frac { \pi \left( \mathcal { M } _ { i } \right) } { \pi \left( \mathcal { M } _ { j } \right) } = \mathcal { B } _ { i j } \frac { \pi \left( \mathcal { M } _ { i } \right) } { \pi \left( \mathcal { M } _ { j } \right) }
\end{equation}

comparing the evidence for both models. For our purpose, we set both model priors $\pi(\mathcal{M})=1/2$, as we give both models equal probability for a given data set, and therefore the odds ratio equals the Bayes factor with $O_{ij}=\mathcal{B}_{ij}$. Using the empirical scale of strength shown in Table \ref{Tab:Evidence}, we can then determine which model is favored.

\subsection{The models}\label{model}
As described in Sec.~\ref{sec:Bayes_theorem_model_comp}, we need two distinct models that are applied to the PSD. For this we use two variants of the background model, based on those presented by \cite{Mathur2011b}, \cite{Karoff2013} and \cite{Kallinger2014}. We call these two models the oscillation model $P_\mathrm{osc}(\nu)$ and noise model $P_\mathrm{noise}(\nu)$.

 The oscillation model $P_\mathrm{osc}(\nu)$ can be expressed as 
\begin{equation}\label{fbm_model}
    P_\mathrm{osc}(\nu)=\sigma_\text{b}+R(\nu)[B(\nu)+G(\nu)]
\end{equation}
where $\sigma_\text{b}$ is assumed to be a flat noise level and $R(\nu)$ the response function, which takes the sampling rate of observations into account, defined by
\begin{equation}
R ( \nu ) = \operatorname { sinc } ^ { 2 } \left( \frac { \pi \nu } { 2 \nu _ { \mathrm { nyq } } } \right)
\end{equation}
with $\nu_\text{nyq}$ being the Nyquist frequency from either the LC or SC observation modality, defined by half of the sampling rate of the signal. The components related to the granulation activity and long term trend of the star, $B(\nu)$, are expressed as a series of Harvey-like functions, given by
\begin{equation}
\label{eq:harvey}
    B ( \nu ) = \sum _ { i = 1 } ^ { 3 } B_i ( \nu ) = \sum _ { i = 1 } ^ { 3 } \frac { \xi a _ { i } ^ { 2 } / b _ { i } } { 1 + \left( \nu / b _ { i } \right) ^{ 4 } }
\end{equation}
which are defined by the root mean sqare (rms) amplitude of the components $a_i$ and the characteristic frequency $b_i$, where $b_i$ is the frequency at which the power of the component is half than that at zero frequency. The factor $\xi$ is used to normalize $\int _ { 0 } ^ { \infty } ( \xi / b ) / \left[ 1 + ( v / b ) ^ { 4 } \right] \mathrm { d } v = 1$, such that the measured $a_i$ actually corresponds to the rms amplitude measured in the light curves, hence yielding $\xi = 2 \sqrt { 2 } / \pi$.

The oscillation region of a solar like oscillator is characterized by a power excess. Here, we describe the power excess, $G ( \nu )$, by a Gaussian function:
\begin{equation}
\label{eq:gaussian}
    G ( \nu ) = H _ { \mathrm { osc } } \exp \left[ - \frac { \left( \nu - \nu _ { \max } \right) ^ { 2 } } { 2 \sigma _ { \mathrm { env } } ^ { 2 } } \right]
\end{equation}
where $H _ { \mathrm { osc } }$, $\nu _ { \max }$ and $\sigma _ { \mathrm { env } }$ are the height, central frequency and standard deviation of the Gaussian corresponding to the power excess.

The second model in this work, $P_\mathrm{noise}(\nu)$, is a variant of the oscillation model $P_\mathrm{osc}(\nu)$. In this model we remove the power excess, leaving us with the equation
\begin{equation}
  \label{eq:model_equation}
  \begin{aligned}
    P_\mathrm{noise}(\nu)&=\sigma_\text{b}+R(\nu)B_3(\nu)
  \end{aligned}
\end{equation}
giving us ten free parameters for $P_\mathrm{osc}(\nu)$ and seven for $P_\mathrm{noise}(\nu)$. We note that many components of our models often exhibit strong correlations that may arise among several of the model free parameters. For this reason, we choose to apply a Bayesian approach based on the nested sampling algorithm \citep{Skilling2006}, which is suited for finding solutions for parameter-degenerate cases.

The next step is to determine priors for the free parameters of each background model by using the PSD of a given object and then feed these priors into \diamonds, a Bayesian Nested sampling code, which will then allow us to determine both the parameter estimates of a given model as well as the corresponding Bayesian evidence. The Bayesian evidence will thus be used to perform model comparison. 
The following two subsections provide the details on the determination of the priors for \diamonds.

\subsection{Determination of priors}\label{sec:det_priors}
An inherent part of every Bayesian approach is the selection of good priors for the analysis, and considerable care should be applied. In this section we describe the process of finding priors for the model $P_\mathrm{osc}$. This is done by computing estimates of all ten free parameters, and by creating prior distributions from those estimates.

\subsubsection{Estimation of \texorpdfstring{$\numax$}{numax}}\label{sec:nu_max_determination}

\begin{figure}
    \includegraphics[width=\linewidth]{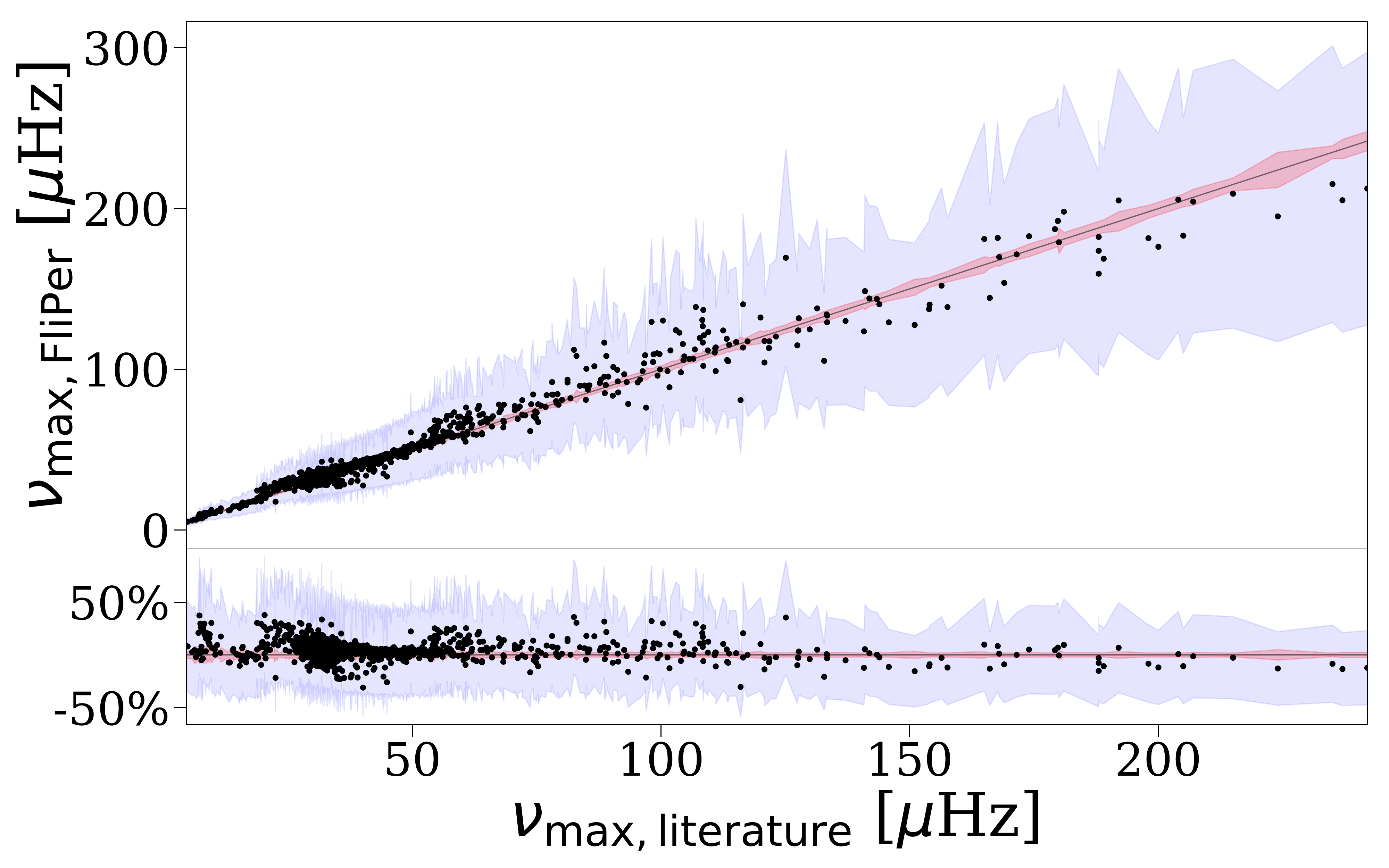}
    \caption{Upper panel: values for $\numax$ extracted from the FliPer method as compared to the values from the APOKASC catalogue. Lower panel: The corresponding residuals in percentile. The red shaded area represents the uncertainties in the catalogue. The blue shaded areas define our $1$-$\sigma$ interval, centered around $\numax$ from FliPer.}
    \label{fig:FliPer_values}
\end{figure}

\begin{figure}
    \includegraphics[width=\linewidth]{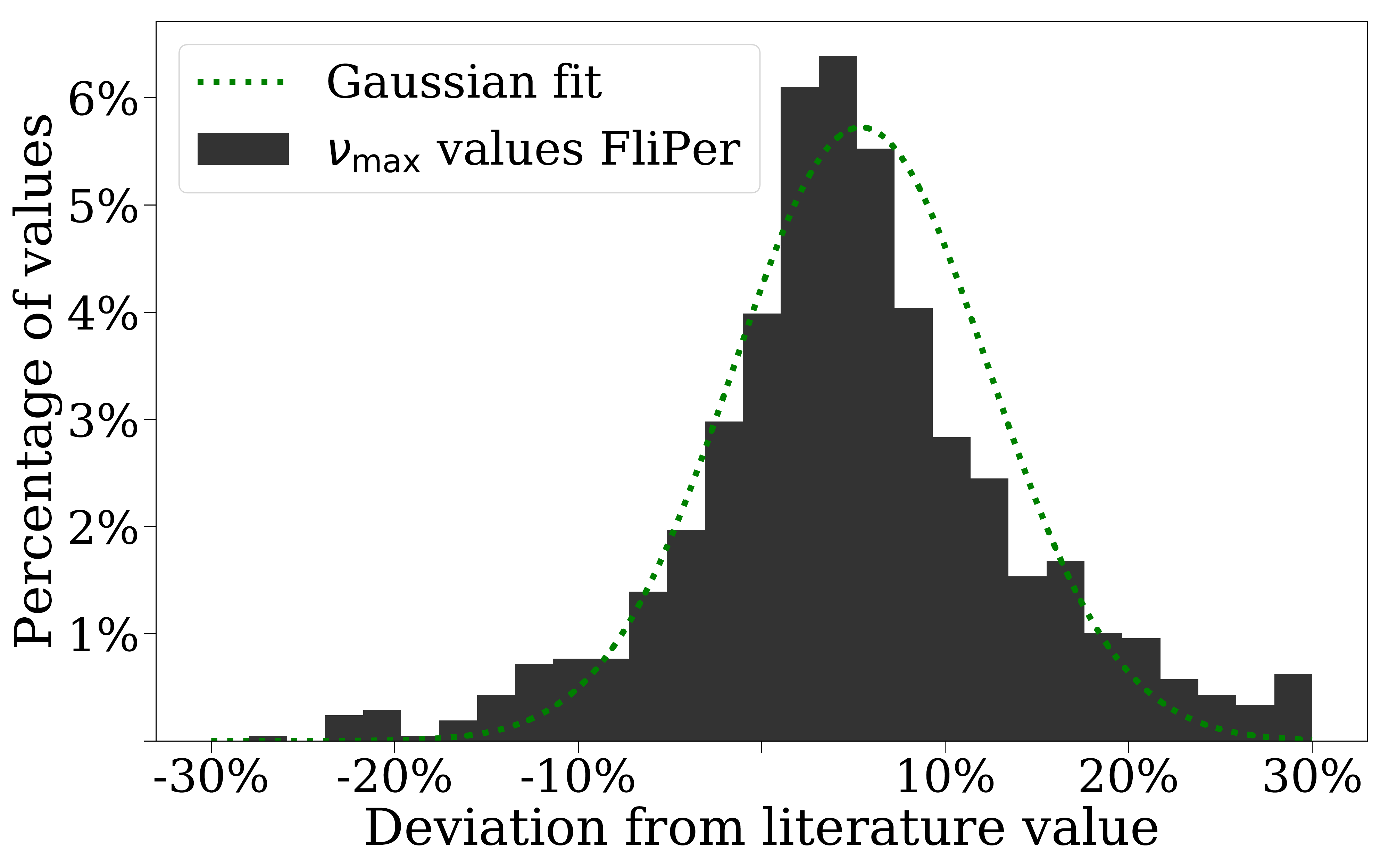}
    \caption{Distribution of values for $\nu_\text{max}$ extracted from the FliPer method. The Y axis shows the percentage of values within a given bin of deviation from a corresponding literature value, while X axis shows the deviation from the literature value in percent. The distribution can be represented by a Gaussian function, whose fit to the histogram is represented by a dotted green line. The mean for this distribution is at $4.95\%$, with a standard deviation of $6.93\%$. For visualization purposes, deviations above 30\% and below -30\% are excluded.}
    \label{fig:FliPer_distribution}
\end{figure}
The most important free parameter in the model $P_\mathrm{osc}$ described in Sect.~\ref{model} is the frequency of maximum oscillation power, $\numax$. This is a characteristic value of a given solar-like oscillator, from which most of the other free parameters can be directly inferred using relatively simple relations (e.g. \citealt{Stello2008,Kallinger2010,Kallinger2014,Pande2018}) which can in turn be used to set up their prior distributions.

The estimation algorithm for $\numax$ used in the \apollo\, pipeline needs to fulfill the following criteria:

\begin{itemize}
    \item It must be computationally inexpensive both for SC and LC data.
    \item It must provide reasonable accuracy, in the sense that the model fitting process described in Sect.~\ref{sec:bg_fitting} is able to find the correct value for $\numax$.
    \item It needs to work for all evolutionary stages of solar-like oscillators.
    \item It must be able to estimate $\numax$ even if the power excess is not clearly  distinguishable.
\end{itemize}

There are multiple approaches to estimate $\numax$ for a given star. Most of them rely on the correlation between the granulation properties of the star and $\numax$. For example \cite{Bastien2013}, using their Flicker method, showed that it is possible to infer $\numax$ from the variation of the stellar flux in the light curve. This method, however, is limited to objects between $2.5 < \log{g} < 4.6\;\mathrm{dex}$, restricting the range of stars that could be analyzed with our approach. 

The method was further improved by \cite{Kallinger2016} who used an iterative approach based on an auto-correlation of the signal. We originally adopted this method for obtaining a good estimate of $\numax$. Nonetheless, we have experienced that it comes with some drawbacks, especially for SC observations with a long baseline. This is because the auto-correlation of a light curve having on the order of $10^5$ data bins, as in the case of \kepler\ SC data covering 3-4 years of observation, is computationally expensive, hence significantly slowing down our pipeline. 

Another potential method for estimating $\numax$ is that presented by \cite{Bell2019}, which estimates $\numax$ using the coefficients of variation. This method mimics the process of visually inspecting a PSD and locating $\numax$, by finding local excesses in power. However, this process only works where the power excess is distinguishable from the background and is the reason why this method is not ideal for our approach, as our method should be also applicable to noisy data, where the power excess cannot be clearly distinguished from the noise. 

Finally, \cite{Bugnet2018} proposed the FliPer method, which is analogous to the Flicker method but applied directly to the PSD of the star. This method is computationally inexpensive, it provides reasonable values for $\numax$ even in conditions of low signal-to-noise in the data, and it works in all evolutionary stages of solar-like oscillators. We therefore choose to adopt it for this work. An exhaustive description of the method can be found in \cite{Bugnet2018}. In summary, the FliPer metric links the variability from all timescales to the surface gravity, $\log(g)$, which in turn is linked to $\nu_\text{max}$. This metric is given by
\begin{equation}
    F _ { p } = \overline { \mathrm { PSD } } - P _ { n }
\end{equation}
where $\overline { \mathrm { PSD } }$ is the mean of the power spectral density and $P _ { n }$ is the photon noise. $F _ { p }$  is determined by a combination of granulation and oscillation power, both heavily dependent on the evolutionary stage of the star, as well as on its rotation.

The computation of the FliPer metric is only applied to LC data. If the pipeline is applied to a SC light curve, it will automatically rebin this light curve to fit into a LC light curve, which is then used to compute the FliPer. Further, the computation of the photon noise in the original work by \cite{Bugnet2018} was done using $K_p$ and the empirical relation in \cite{Jenkins2010}. In this work we make use of the adpated method from \cite{Bugnet2019} to compute this background noise $P_n$. This is done by averaging the last 100 data points adjacent to the Nyquist frequency. While the original method from \cite{Bugnet2018} yields a value for $\numax$ closer to the literature values from the APOKASC catalogue, this approach is also useful if there is no \kepler\, magnitude available for a given star.

The resulting values for $\nu_\text{max,FliPer}$ are shown in Fig.~\ref{fig:FliPer_values}. The FliPer metric provides us with reasonable values of $\numax$, hence allowing for adequately setting up priors for the remaining free parameters of the background models. For incorporating cases where the estimated $\numax$ may not be very accurate, we consider an adequate search range for the fit, as the blue shaded regions indicate. These regions (representing the $1$-$\sigma$ range where we expect $\numax$ using FliPer, given the standard deviation of our prior for $\numax$) do all include the corresponding values from the APOKASC catalogue.

In general however, the FliPer values alone already give us a good estimate of $\numax$, as shown in the distribution of Fig~\ref{fig:FliPer_distribution}. This distribution shows that the FliPer value for $\nu_\text{max}$ exhibits a systematic deviation of $4.95$\,\% and a random error within $1$-$\sigma$ of $6.93$\,\% with respect to the values reported in the literature.

\subsubsection{Determination of priors for the full background model}\label{sec:prior_ranges}

\begin{figure*}[!htb]
    \includegraphics[width=\textwidth]{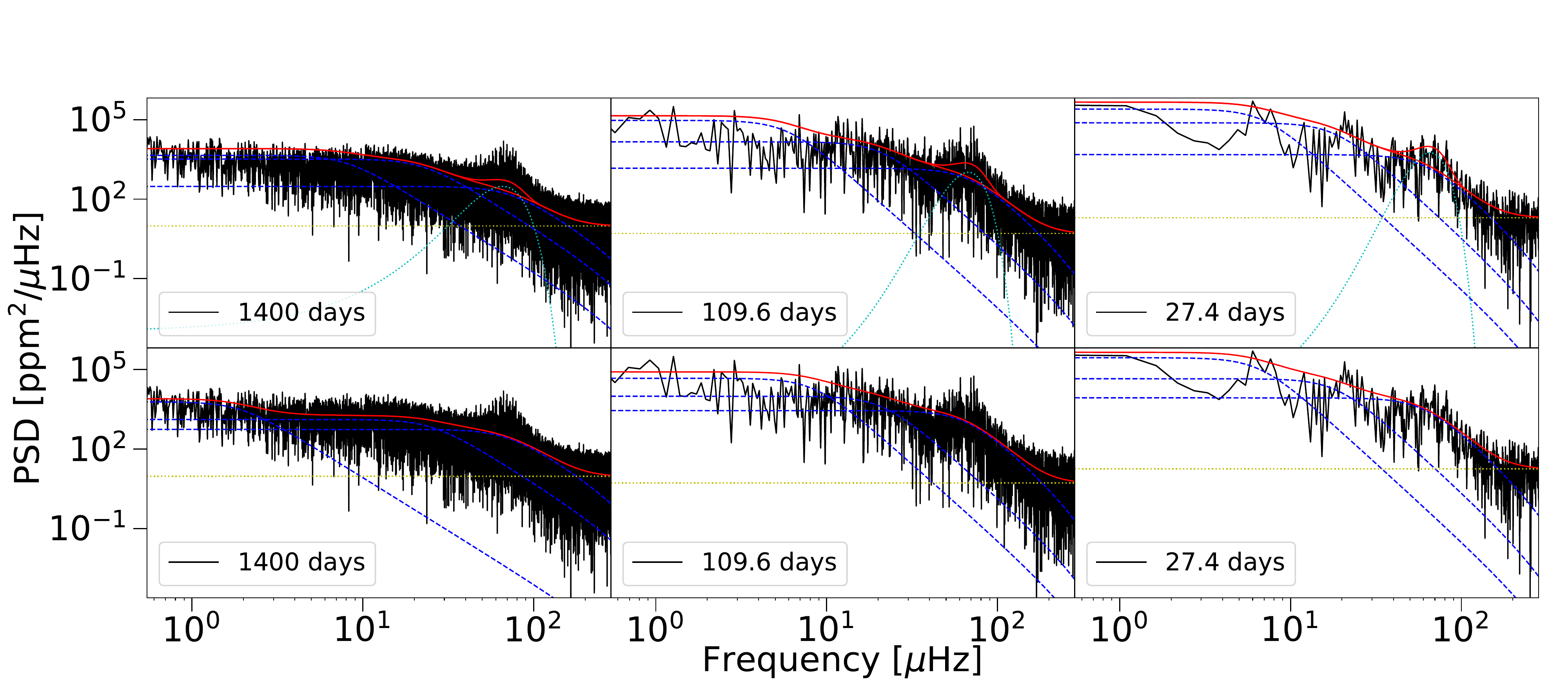}
    \caption{The fit provided by \diamonds\,\,for three different observation lengths for KIC 9903915. The upper row shows the full background model, the lower row shows the noise only model. Different observation lengths are split into three columns: The first column shows the result for the full 1400 day light curve, the second for 109.6 days, the third for 27.4 days. We find high evidence for all three observation lengths: $\log O = 990.53$ for 1400 days, $\log O = 75.42$ and $\log O = 13.86$ for 27 days. Even at 27 days we are still above the threshold of  $\log O = 5$ for strong evidence.}
    \label{fig:Bayes_value_example}
\end{figure*}
The parameters $a_2$, $a_3$, $b_2$ and $b_3$, that define two of our three Harvey like functions in Eq.~(\ref{eq:harvey}), are estimated using scaling relations from \cite{Kallinger2014}, which relate these parameters directly to $\nu_\text{max}$. For our first Harvey like function, representing the long trend variation, we adapt the following: $a_1$ is estimated using the same scaling relations as $a_2$ and $a_3$, the prior distribution for $b_1$ is directly determined and not calculated (see Tables \ref{Tab:Parameter_ranges} and \ref{Tab:Parameter_ranges_dwarfs}). The flat background noise $$\sigma_\text{b}$$ is determined by computing the median of the adjacent 100 points to the Nyquist frequency, and converting it to the mean using the relation between median and mean for the $\chi^2$ distribution with two degrees of freedom \citep{Appourchaux2014}. These seven parameters ($a_1$, $a_2$, $a_3$, $b_1$, $b_2$, $b_3$ and $\sigma_\text{b}$) fully estimate $P_\mathrm{noise}(\nu)$ and are also used for the Harvey like function and background noise in $P_\mathrm{osc}(\nu)$. The remaining free parameters for which an initial guess has to be obtained are $H_{\mathrm{osc}}$ and $\sigma_\text{env}$, which are related to the Gaussian envelope presented in Eq.~(\ref{eq:gaussian}). The parameter $\Gamma_\text{env}$, namely the full width at half maximum of the Gaussian envelope, is determined using \citep{Mosser2012}
\begin{equation}
    \Gamma_\text{env}=0.66\nu_\text{max}^{0.88}
\end{equation}
and it is related to $\sigma_\text{env}$ with 
\begin{equation}
    \sigma_\text{env}=\frac{\Gamma_\text{env}}{2\sqrt{2\log 2}}
\end{equation}
An estimation for $H _ { \mathrm { osc } }$ can instead be found by using $\numax$ and $\sigma_\text{env}$ to restrict the PSD to the region $\numax - \sigma_\text{env} < \nu < \numax + \sigma_\text{env}$, hence computing the median in this region to get our estimate of $H _ { \mathrm { osc } }$. This concludes our set of ten parameters for $P_\mathrm{osc}(\nu)$.

We have in principle multiple options to use these input guesses for the free parameters of the  models when constructing priors because it is possible to adopt different prior distributions in a Bayesian inference process. Generally we choose to use flat (i.e. uniform) priors for all parameters except $\nu_\text{max}$. Even if we have an estimate for a central value for each of the free parameters as described above, we still prefer to adopt uniform priors for most of them because some of the parameters exhibit an interval of variation that could be well constrained from scaling relations, and because uniform priors are computationally the most efficient ones. For $\nu_\text{max}$ instead, it is more critical to make sure that the actual value of this parameter is properly fitted during the Bayesian inference. This is because our purpose is that of detecting the presence of a possible power excess in an automated way, so that identifying the correct location of the power excess is a mandatory step to make the whole analysis successful. The prior knowledge through our analysis of the FliPer metrics, which takes into account the systematic and random errors in the FliPer as seen in Fig.~\ref{fig:FliPer_distribution}, suggests that a normal distribution represents an adequate choice for the prior of this parameter. This has the additional advantage that we further mitigate fluctuations in our estimation of $\nu_\text{max}$. This is because normally distributed priors are not as restrictive as flat priors, meaning that  there is a non-zero prior value along the full range of the given free parameter. Flat priors in contrast, assign an equal prior probability within a given range and a sharp transition to zero prior probability outside of the prior range, thus cutting out any possibility of detecting the actual value of the free parameter in case this is falling outside the prior boundaries.

\begin{table}
\caption{Prior ranges for our red giant sample. }           
\label{Tab:Parameter_ranges}      
\centering                                      
\resizebox{\columnwidth}{!}{%
\begin{tabular}{c|c c c c c c c c}          
\hline\hline                      
&\small{$\sigma_\text{b}$}& \small{$a_i$}&\small{$b_1$}&\small{$b_2$}&\small{$b_3$}&\small{$H _ { \mathrm { osc } }$}& \small{$\nu_\text{max}$}& \small{$\sigma_\text{env}$}\\ 
\hline                                   
\begin{tabular}[c]{@{}l@{}}Low\\Up\end{tabular} & 
\begin{tabular}[c]{@{}l@{}}0.5\\2\end{tabular} & 
\begin{tabular}[c]{@{}l@{}}0.05\\$\max(\text{PSD})$\end{tabular} & 
\begin{tabular}[c]{@{}l@{}}$\nu_\text{min}$\\$b_{2,min}$\end{tabular} & 
\begin{tabular}[c]{@{}l@{}}0.7\\1.3\end{tabular}& 
\begin{tabular}[c]{@{}l@{}}0.7\\1.3\end{tabular}& 
\begin{tabular}[c]{@{}l@{}}0.1\\3.5\end{tabular}&
\begin{tabular}[c]{@{}l@{}}0.7\\1.3\end{tabular}&
\begin{tabular}[c]{@{}l@{}}0.7\\1.3\end{tabular}  \\
\hline                                             
\end{tabular}
}
\tablefoot{Up and low values are variations in relative units of the centroid values determined in Sect.~\ref{sec:det_priors}. The ranges for $\numax$ are the $1$-$\sigma$ values of the normal distribution, with the parameter determined in  as its mean value. The low value for $b_1$ equals the smallest frequency in the power spectrum, the up value always the lower end of $b_2$ (i.e. $0.7*b_2$). The upper limit of $a_i$ equals the maximum power in the PSD.}
\end{table}

\begin{table}
\caption{Same as Table \ref{Tab:Parameter_ranges} but for our MS and subgiant sample.}             
\label{Tab:Parameter_ranges_dwarfs}      
\centering                                      
\resizebox{\columnwidth}{!}{%
\begin{tabular}{c|c c c c c c c c}          
\hline\hline                      
&\small{$\sigma_\text{b}$}& \small{$a_i$}&\small{$b_1$}&\small{$b_2$}&\small{$b_3$}&\small{$H _ { \mathrm { osc } }$}& \small{$\nu_\text{max}$}& \small{$\sigma_\text{env}$}\\ 
\hline                                   
\begin{tabular}[c]{@{}l@{}}Low\\Up\end{tabular} & 
\begin{tabular}[c]{@{}l@{}}0.5\\3\end{tabular} & 
\begin{tabular}[c]{@{}l@{}}0.01\\$\max(\text{PSD})$\end{tabular} & 
\begin{tabular}[c]{@{}l@{}}$\nu_\text{min}$\\$b_{2,min}$\end{tabular} & 
\begin{tabular}[c]{@{}l@{}}0.7\\1.3\end{tabular}& 
\begin{tabular}[c]{@{}l@{}}0.7\\1.4\end{tabular}& 
\begin{tabular}[c]{@{}l@{}}0.1\\3.5\end{tabular}&
\begin{tabular}[c]{@{}l@{}}0.8\\1.2\end{tabular}&
\begin{tabular}[c]{@{}l@{}}0.7\\2\end{tabular}  \\
\hline                                             
\end{tabular}
}
\end{table}

\subsection{Background fitting and \diamonds}\label{sec:bg_fitting}
To fit the models $P_\mathrm{osc}$ and $P_\mathrm{noise}$ to a given PSD, we apply the Bayesian parameter estimation tool \diamonds\footnote{https://github.com/EnricoCorsaro/DIAMONDS} \citep{Corsaro2015}. \diamonds\, applies Bayes' theorem, as described in Sec.~\ref{sec:Bayes_theorem_model_comp}. Eq.~(\ref{eq:evidence_equation}) is a multi-dimensional integral and as the number of dimensions increases, solving it becomes very hard both analytically and numerically. \diamonds\, overcomes this problem by applying the Nested sampling Monte Carlo \citep[NSMC;][]{Skilling2006} algorithm. This algorithm both samples the posterior probability distribution as well as the Bayesian evidence, which in the context of this work is used to compare the models $P_\mathrm{osc}$ and $P_\mathrm{noise}$ and simplifies the multi-dimensional problem of Eq.~(\ref{eq:evidence_equation}) into a one dimensional problem. The simplification is done by considering the \textit{prior mass} $\mathrm { d } X = \pi ( \boldsymbol { \theta } | \mathcal { M } ) \mathrm { d } \boldsymbol { \theta }$ such that
\begin{equation}
    X \left( \mathcal { L } ^ { * } \right) = \int _ { \mathcal { L } ( \theta ) > \mathcal { L } ^ { * } } \pi ( \boldsymbol { \theta } | \mathcal { M } ) \mathrm { d } \boldsymbol { \theta }
\end{equation}
 defining a fraction of volume under the prior PDF constrained by $\mathcal { L } ^ { * }$, with $\mathcal { L } ^ { * }$ being some fixed value of the likelihood. This therefore reduces to a one-dimensional integral with
\begin{equation}\label{eq:evidence_eq}
    E  = \int _ { 0 } ^ { 1 } \mathcal { L } ( X ) \mathrm { d } X \, .
\end{equation}
Assuming that one has $N_{ \text { nest } }$ pairs of $\left\{ \mathcal { L } _ { i } ^ { * } , X _ { i } \right\}$, Eq.~(\ref{eq:evidence_eq}) can be rewritten as
\begin{equation}\label{eq:evidence_eq_discrete}
    E = \sum _ { i = 0 } ^ { N _ { \text { nest } } - 1 } \mathcal { L } _ { i } ^ { * } w _ { i } \, .
\end{equation}
In practice, this is solved by iteratively setting a likelihood constraint that is higher than a previous one for each iteration, peeling of a thin shell of prior mass and evaluating the evidence using Eq.~(\ref{eq:evidence_eq_discrete}). Nevertheless, the NSMC algorithm can quickly become computationally expensive because of the increasing difficulty in obtaining new good sampling points at each iteration. \diamonds\, overcomes this problem by applying simultaneous ellipsoidal sampling (SES) \citep{Feroz2008,Shaw2007} based on a preliminary clustering of the set of live points at a given iteration. 

\diamonds\, also allows to choose different prior distributions for different parameters of our models. This is specifically useful for our approach and not a standard feature in other existing codes of this kind. The parameter ranges for all free parameters are given in Table \ref{Tab:Parameter_ranges} for LC data and in Table \ref{Tab:Parameter_ranges_dwarfs} ranges for SC data. These should be interpreted as variations around the central values determined in Sect.~\ref{sec:prior_ranges} and are made small enough to allow \diamonds\, to efficiently converge to a solution, but are also large enough to properly account for uncertainties in the free parameters. For the amplitudes $a_i$ and $H_\mathrm{osc}$ a large range was given deliberately due to the character of the fitting process and due to the large variations in the parameter value and dependencies on other parameters. A special case is the prior $b_1$, the cutoff frequency of the first Harvey component. The parameter $b_1$ is not directly inferred from the PSD or from $\numax$. For $b_1$ to be appropriate, it always has to fulfill $\nu_\text{min}<b_1<\min(b_2)$, where $\nu_{min}$ equals the smallest frequency in the power spectrum and $\min(b_2)$ the lower end of the prior for $b_2$.

The priors described in the previous sections are used by \diamonds\, to fit both models $P_\mathrm{osc}$ and $P_\mathrm{noise}$ to the power spectral density of a given star. The estimated parameters, including their marginal distributions, as well as the Bayesian evidences $E_{osc}$ and $E_{noise}$, for models $P_\mathrm{osc}$ and $P_\mathrm{noise}$ respectively, are all outputs of the fitting process. Using their respective evidence, we compute the odds ratio of the models $O_{osc,noise}$ and using the empirical scale of strength shown in Table \ref{Tab:Evidence} we determine if the oscillation model $P_\mathrm{osc}$ is to be favoured over the noise model $P_\mathrm{noise}$, giving us an indication if there is an oscillation signal for a given data set.

This is especially useful for power spectra where a power excess is not clearly distinguishable by eye. One example, that shows the potential of this approach is illustrated in Fig.~\ref{fig:Bayes_value_example}. In this example the method is applied to a test star from our sample of red giants, using an observation length of $\SI{1400}{days}$, $\SI{109.6}{days}$ and $\SI{27.4}{days}$. It appears that in the case of $\SI{27.4}{days}$, we cannot clearly discern the oscillation region anymore in the PSD. If the atmospheric properties of the star, such as surface gravity and effective temperature, are available, it is possible to predict the frequency position of a power excess in the region where it is actually found. In real applications with similar signal-to-noise ratio as in the example shown in Fig.~\ref{fig:Bayes_value_example}, visual inspection alone is insufficient to claim a detection.

\subsection{Determination of \texorpdfstring{$\dnu$}{dnu}}
The pipeline also determines the large frequency separation $\dnu$ if the fit of the background model is successfully applied. For this, we no longer rely on a Bayesian approach, but use classical means. We subtract the background without the Gaussian envelope from the PSD of the star and restrict the resulting spectrum to the oscillation region through the determined values for $\numax$ and $\sigma_\text{env}$. This region is first slightly smoothed and then auto-correlated, allowing us to detect the typical regularity in the frequency pattern of the $p$ modes. Through the well known relation between $\numax$ and $\dnu$ \citep{Stello2009} we can compute an initial guess for $\dnu$ using
\begin{equation}
    \Delta\nu = 0.259 \nu_\text{max}^{0.765}
\end{equation}
by \citet{Huber2011}.
Using this value as a reference, we can locate the closest peak that is obtained from the auto-correlation. We then fit a normal distribution to this peak and take its resulting mean as the estimated $\dnu$.

\section{Implementation of the \texorpdfstring{\apollo\,}{apollo} pipeline}\label{sec:technical_details}
\apollo\, is a fully integrated Bayesian fitting pipeline. Its core is written in Python, with extensive use of highly optimized scientific libraries \citep{scipy,numpy,astropy:2013,astropy:2018}. It also makes direct use of \diamonds, for which we created both a Python interface as well as direct parsing of the binary output. The pipeline also generates the necessary run files for \diamonds\, and analyzes its outputs. It possesses full multi-processing capabilities, allowing for the analysis of multiple stars in parallel when modern multi-core processors are in use. This allows for quick evaluations of large data sets: Our sample of red giants, consisting of 1071 objects with the majority having four years of observation and consisting of around 68\,000 data points, takes six hours to complete using the AMD Threadripper architecture with all 32 threads in use. Extrapolating from this, the average run time for one object, including the fit for both models, takes around 20\,s. In comparison, performing the analysis sequentially (one star at a time on one thread) takes between five and ten minutes per star.

The pipeline is also highly configurable. It uses the JSON format as input and output files, which both allows machine readability as well as the possibility for humans to read these files and interpret their content. The input files contain various configuration flags, making the pipeline very flexible for various usages. The output files contain all results from \diamonds, as well as priors, estimations of parameters and a myriad of other data, that can be used to further analyze the result.

The full code as well as documentation and examples are available at the github repository\footnote{https://github.com/MarcoMuellner/APOLLO}. All further improvements and changes will also be available there.

\section{Calibration}
\subsection{Reliability of the method}
The first step in the analysis using \apollo\, is to compare the results obtained with the pipeline to the results from the APOKASC and LEGACY samples, both consisting of stars observed with \kepler, that we use in this work to check the robustness of the method. We also introduce the term \emph{completion rate}, which is the number of completed runs for a given sample of stars. In order for a run to qualify as completed, the following four criteria must be fulfilled:

\begin{enumerate}
    \item The estimations of the free parameters must have been successfully carried out by \diamonds, meaning that the code was able to perform a fit for both background  models without errors.
    \item The Bayes factor $O_\text{osc,noise}$ as defined in equation \ref{eq:bayes_factor} must exceed the criterion of strong significance, which is $O_\text{osc,noise} > 5$.
    \item Runs that result in a $\nu_\text{max}$ outside of $\SI{1}{\sigma}$ of the values from APOKASC and Legacy catalogs are manually checked and discarded if their fit is not correct. If this is not the case, the results are kept, even if $\numax$ is outside of $\SI{1}{\sigma}$ of the catalogue values. This step is only necessary for the calibration of the method.
    \item Power spectra that show contamination by binaries \\\citep{Colman2017} are also discarded, if the pipeline does not detect the oscillation region correctly, but rather the excess due to contamination.
\end{enumerate}

Following these criteria, we get a completion rate of $\SI{95.5}{\%}$ for our red giant sample, completing 1023 out of 1071 objects. The reasons for discarding 48 objects out of the 1071 analyzed are twofold:
\begin{itemize}
    \item 36 objects show a wrong estimate of $\numax$ from the FliPer metric. This causes \diamonds\, not to be able to adequately sample the actual $\numax$ values since the normal prior centroid is far from the proper solution.
    \item 12 objects contain contamination of binaries, which leads to \diamonds\, detecting the excess of the contamination as the oscillation region.
\end{itemize}

\begin{figure}[!htb]
    \centering
    \includegraphics[width=\linewidth]{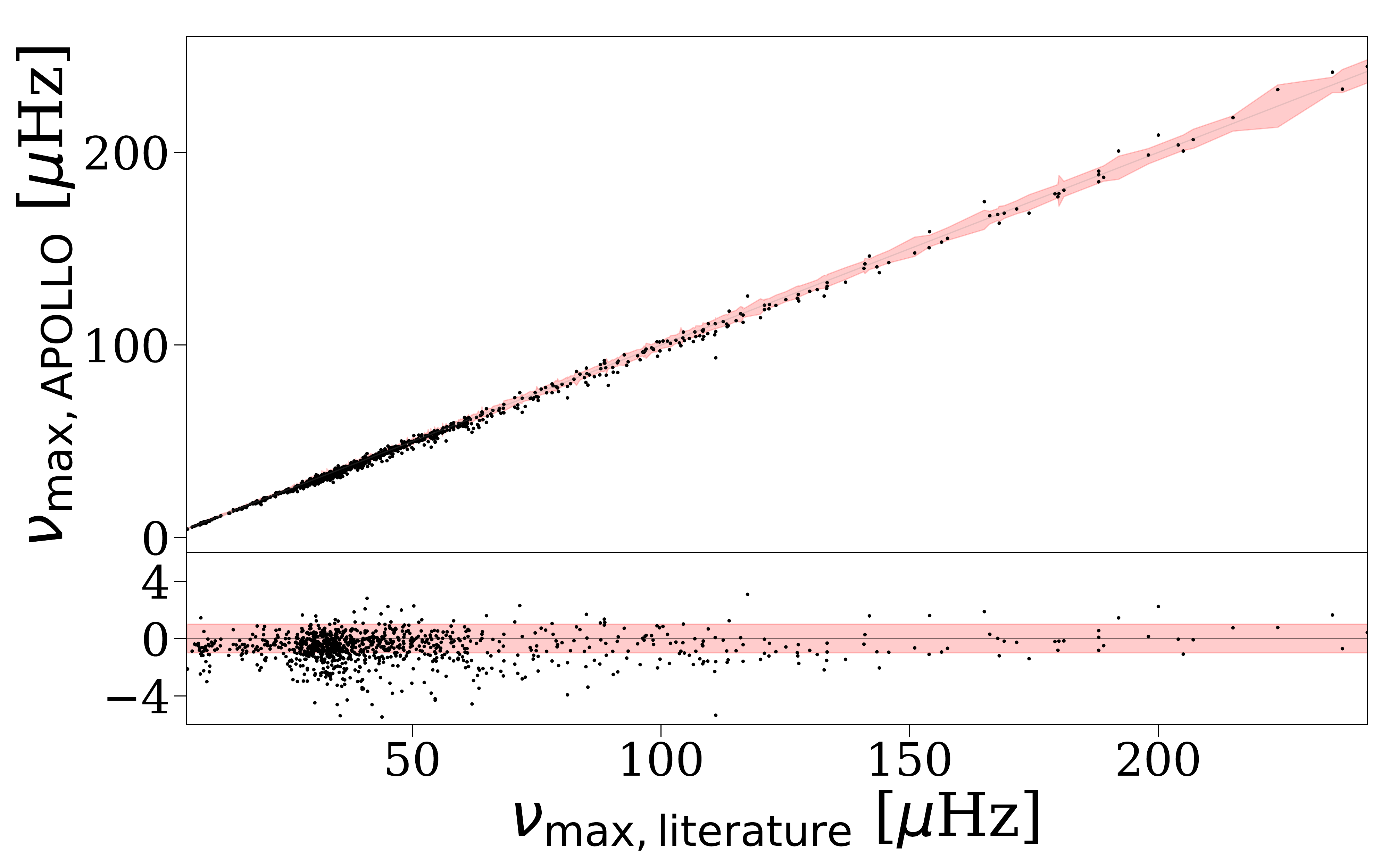}
    \caption{Resulting $\nu_\text{max,\apollo}$ values of the \apollo\, pipeline in relation to catalogue values of our red giant sample. The upper panel shows the comparison of $\nu_\text{max,\apollo\,}$ to the literature, with the red shaded area marking the $1$-$\sigma$ uncertainties in the literature. In the lower plot we show the residuals in sigma of the literature value and again mark the uncertainties in the red shaded area.}
    \label{fig:nu_max_values}
\end{figure}

\begin{figure}[!htb]
    \centering 
    \includegraphics[width=\linewidth]{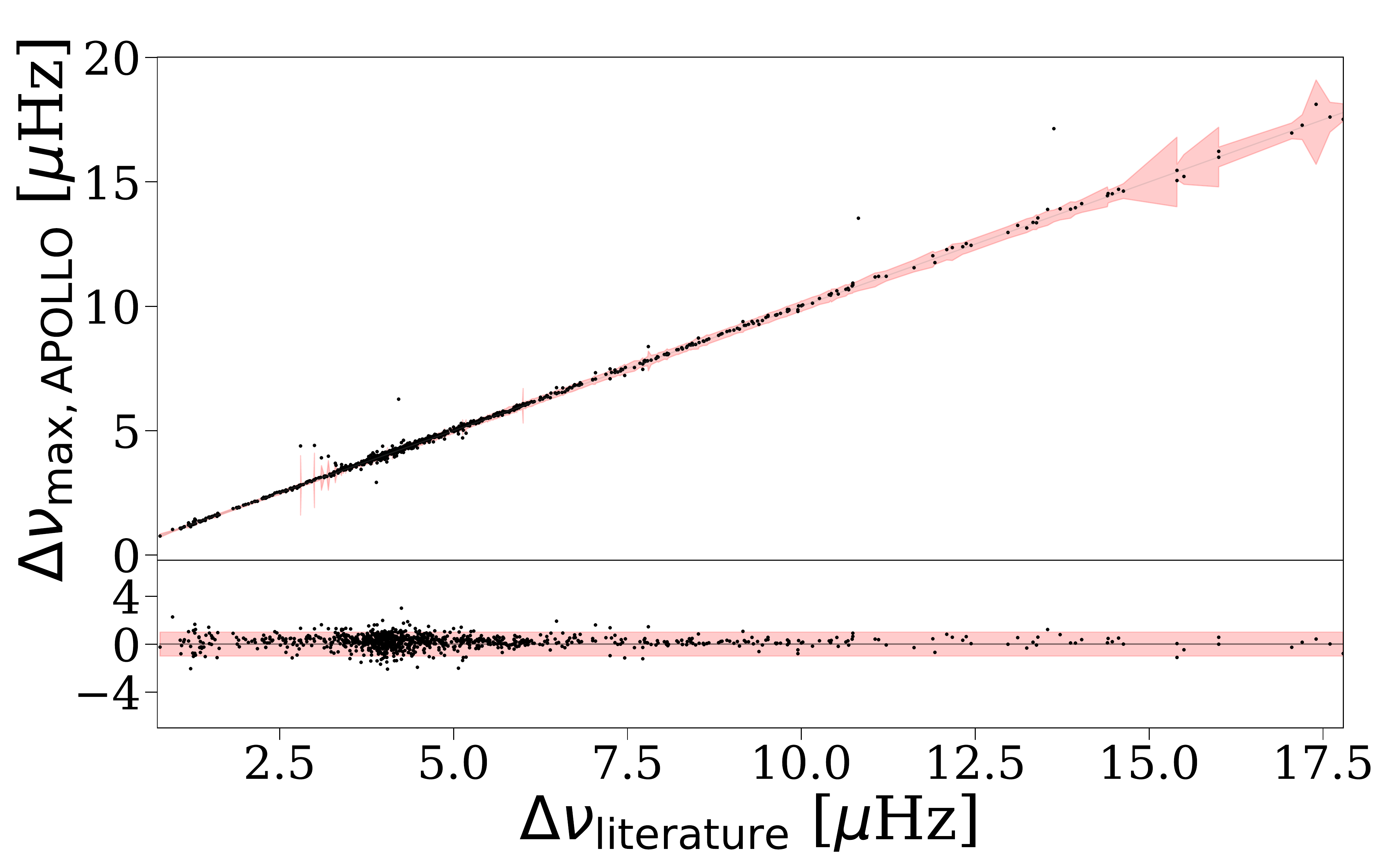}
    \caption{Similar as Fig.~\ref{fig:nu_max_values}, but for the large frequency separation $\dnu$.}
    \label{fig:delta_nu_values}
\end{figure}

The resulting values from our red giant sample for $\numax$ for the completed 1023 objects are shown in Fig.~\ref{fig:nu_max_values} and their corresponding values for $\dnu$ in Fig.~\ref{fig:delta_nu_values}. Out of these 1023 objects, $65\%$ agree within $1$-$\sigma$ to the catalogue values of $\numax$. For $\dnu$ we find excellent agreement with the literature values, with $91\%$ of all values falling within $1$-$\sigma$ uncertainty from the literature values. 

We performed the same analysis for our MS sample. Initially we applied the same boundaries for our priors for both the MS sample as well as the red giants. The prior distributions for $\numax$ and $\sigma_\mathrm{env}$ were consequently too large for the main sequence sample, making it impossible to reliably fit this sample. We therefore restricted the prior distributions further than we did for the red giants. Our estimation of the background noise $\sigma_b$ turned out to be too small for some of the MS sample, which we mitigated by slightly increasing the upper limit of the prior distribution for $\sigma_b$. 

\begin{figure}[!htb]
    \centering
    \includegraphics[width=\linewidth]{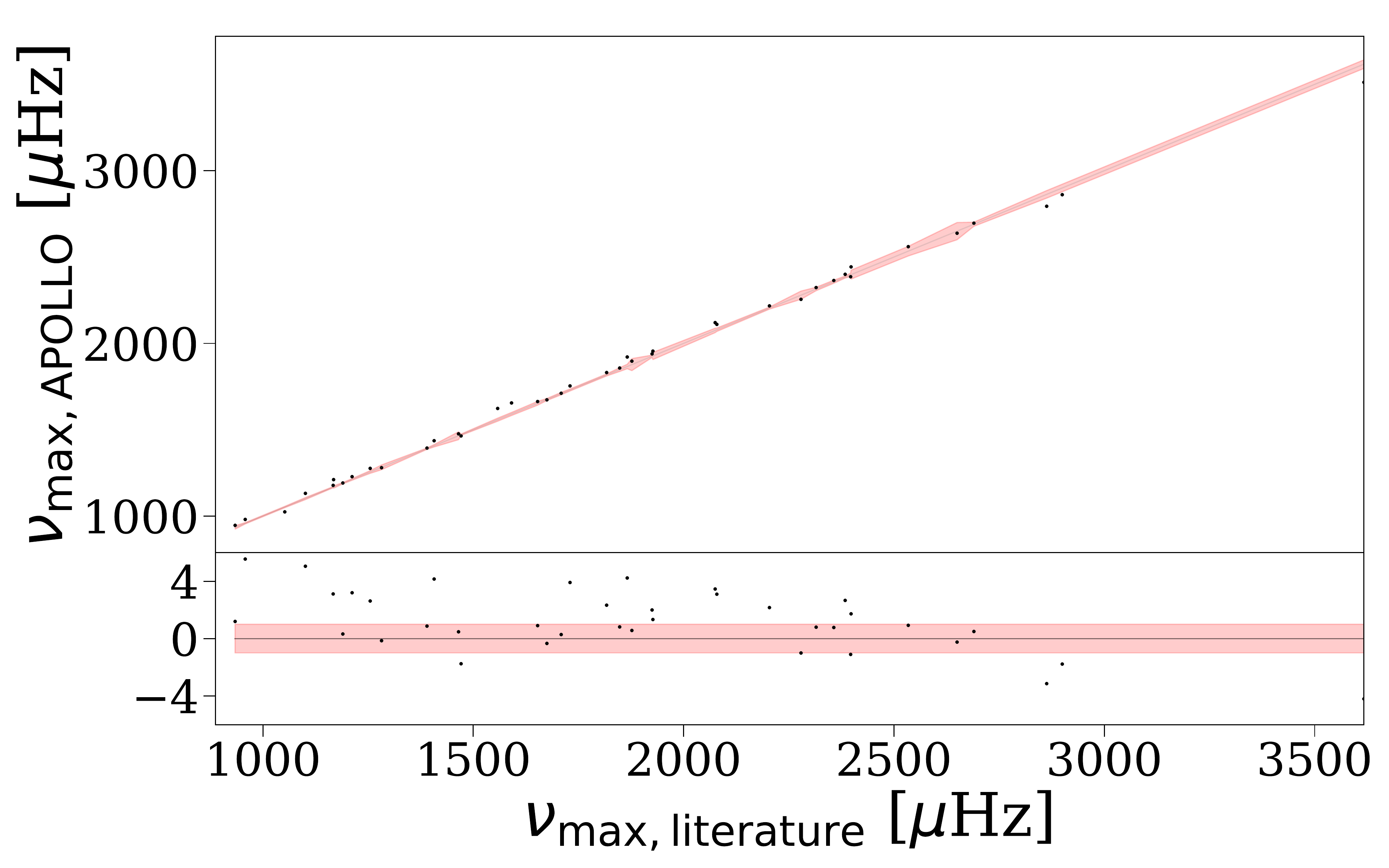}
    \caption{Similar to Fig.~\ref{fig:nu_max_values} but for the sample from the LEGACY catalogue.} 
    \label{fig:nu_max_progression_legacy}
\end{figure}

The resulting values for $\numax$ are shown in Fig.~\ref{fig:nu_max_progression_legacy}. For our dwarfs we find a completion rate of $85\%$. From the nine objects that failed, for three no fit was found, and one object showed a value for $\ln(O_{osc,noise}) < 5$. The remaining five failed due to a wrong estimate of $\numax$ from FliPer. For our main sequence sample $35\%$ agree within $1$-$\sigma$ with the values from the LEGACY catalogue. In percentile, $85\%$ agree within $10\%$, $68\%$ within $5\%$ and $51\%$ within $3\%$ of the literature value. Most likely the explanation for these large deviations in terms of $\sigma$ could be a application of a different background model applied by \cite{Lund2017}, where they only used two Harvey-like functions, as well as a different representation of the oscillation region through a series of Lorentzian functions.


\subsection{Reduction of observation length}\label{obs_length_reduction}
With the advent of the TESS space mission, we are now dealing with data sets that exhibit various observing lengths. In this regard, it is also relevant for us to test the performance of the pipeline in observing conditions different than those of \kepler. To do this, we reduce the baseline of the \kepler\,\,observations to adapt to those covered by TESS, namely $356.2$\,d, $109.6$\,d, $82.2$\,d, $54.8$\,d and $27.4$\,d. We use the full sample of our red giants, and compute fits for all 1023 objects, that were completed using the default baseline,  in all the five observation lengths provided by TESS. The completion rate (from the 1023 objects) yields $100\%$ for $T_{\rm obs} = 356.2$\,d and $99\%$ for observation times $109.6$\,d, $82.2$\,d and $54.8$\,d. For our lowest observation time of $27.4$\,d we find a completion rate of $95\%$, showing that our pipeline is very much insensitive to the observation baseline, making it very suitable for TESS data.

\begin{figure*}[!htb]
    \centering
    \includegraphics[width=\textwidth]{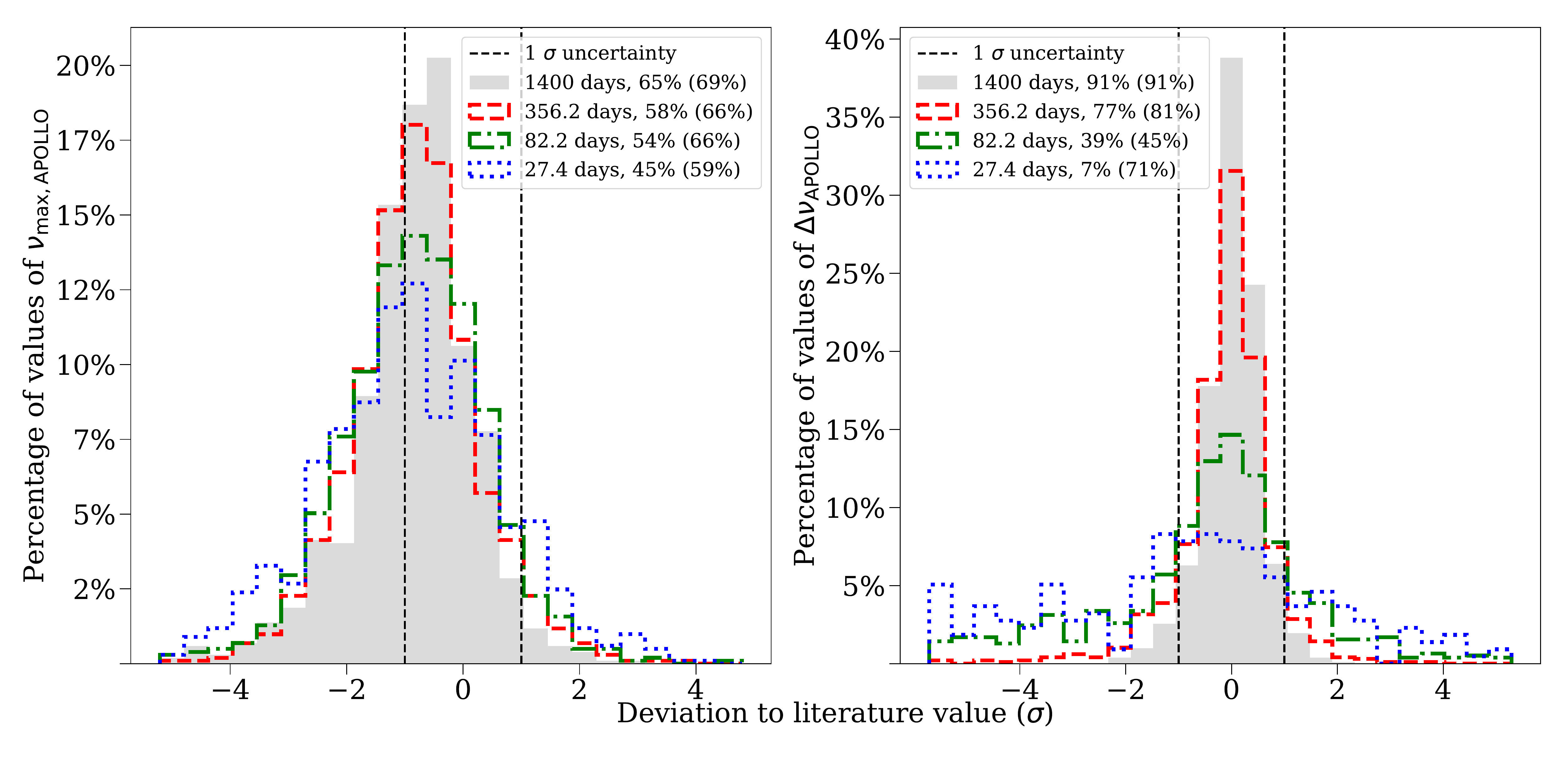}
    \caption{Distribution of values for $\numax$ (left panel) and $\dnu$ (right panel) in different observation lengths. Grey filled bars represent the values for 1400 days of observation, smaller observation lengths are shown in different colors. The red dashed lines represent the $1$-$\sigma$ range of the literature values. The legend states the percentage value of stars within $1$-$\sigma$. In brackets we note the percentage value when taking the uncertainties of the fit into account. For $\numax$, using 27\,d observations $19.88$\,\% of our sample falls outside of the $1$-$\sigma$ range in comparison to the full observation length. For $\dnu$ this decrease is much stronger, with $83.91$\,\% of the sample falling outside of $1$-$\sigma$ using 27\,d of observation.}
    \label{fig:Observation_length_behaviour}
\end{figure*}

Again, we compare the results from these fits to the values for $\numax$ from the APOKASC catalogue. The result for our baseline and the five TESS observation lengths is shown in Fig.~\ref{fig:Observation_length_behaviour} (left panel). The result shows, as expected, a broadening of the distribution for smaller observation lengths. With the lowest observation time of $27.4$\,d, $45\%$ of the values for $\numax$ still agree with the literature values within $1$-$\sigma$. We also find a systematic deviation to the literature values, which can be explained through the choice of a different background formalism.

By comparing the result for $\dnu$ from the pipeline to the catalogue values, the effect of shorter observation lengths is significantly stronger than that of $\numax$. As can be seen by looking at Fig.~\ref{fig:Observation_length_behaviour} (right panel), we find a strong broadening of the distribution, especially for the very short observing lengths. This is an effect of the reduced signal to noise ratio in stars with short observation lengths and amplified by the decreasing resolution of the power spectrum. This is reflected in the total values falling within $1$-$\sigma$ of the catalogue values: For $109.6$\,d already, only $50\%$ of all objects fall inside the uncertainty. Using $27.4$\,d reduces this further to only $7$\,\%. 

\subsection{Detection of fainter stars}
To simulate the expected magnitude range of stars extracted by the TASC photometry pipeline \citep{Handberg2019} for the red giant sample, we add random, normally distributed noise to the light curves of a sample of stars. We compute the white noise from the PSD, by calculating the mean of the preceding 100 data points below the Nyquist frequency. In the following we convert this noise to the corresponding \kepler\, magnitude $Kp$ and relate this to the TESS magnitude $I_C$ with $I_C=Kp-5$ \citep{Stassun2018} using the relation by \cite{Pande2018}.

\begin{figure}[!htb]
    \centering
    \includegraphics[width=\linewidth]{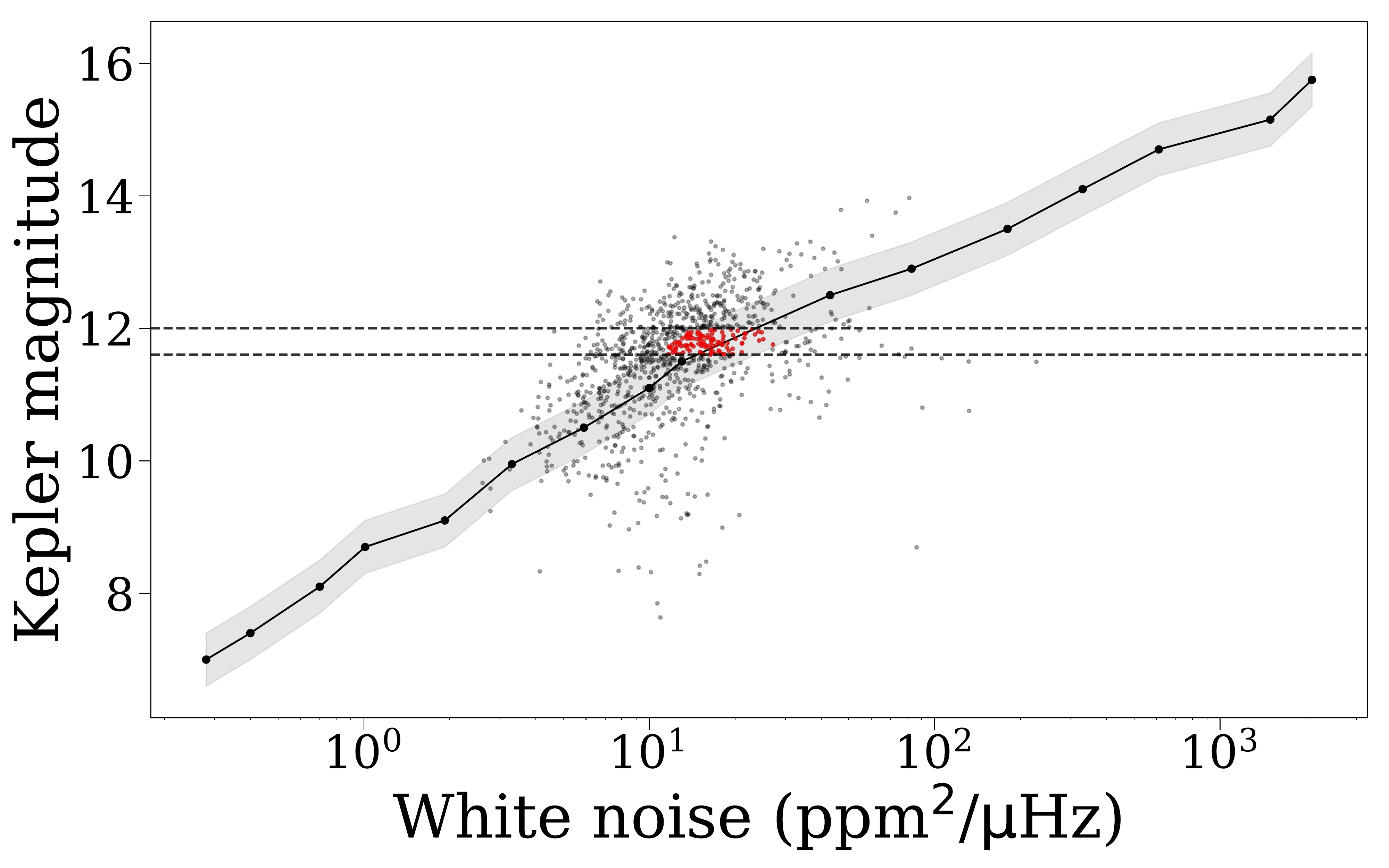}
    \caption{Relation between \kepler\, magnitude $Kp$, and white noise. The black solid line represents the relation from \cite{Pande2018}, the grey shaded area represents the white noise range corresponding to a variation of $0.4$\,mag. Our sample is represented by the grey and red dots, where the red dots constitute the sample used in the magnitude analysis and is identified by the intersection of the horizontal lines to the relation found by \cite{Pande2018}. The horizontal grey dashed lines show the range of these points within $11.6\,\mbox{mag} < Kp < 12\,\mbox{mag}$}
    \label{fig:magnitude_noise_relation}
\end{figure}

The actual sample used for this simulation test is obtained by reducing our initial sample of \kepler\, red giants based on the following conditions:
\begin{itemize}
    \item We restrict ourselves to stars that follow the relation from \cite{Pande2018} within $\SI{0.4}{mag}$, to reduce the uncertainty in the calculated magnitudes from the white noise.
    \item We choose stars that show a magnitude between $11.6\,\mbox{mag}<Kp<12\,\mbox{mag}$. This allows to simulate sufficient points to the faintest objects at $15\,\mbox{mag}$ ($I_C=10\,\mbox{mag}$) and assigns all simulated stars a similar starting value.
    \item All objects must fall within $1$-$\sigma$ of the catalogue values for $\nu_\text{max}$
\end{itemize}
Applying these criteria yields 84 valid objects which are shown in Fig.~\ref{fig:magnitude_noise_relation}. From this sample we compute ten snapshots of white noise level, corresponding to magnitudes between $12\,\mbox{mag}<Kp<15\,\mbox{mag}$, a range that incorporates the upper limit considered in the TASC data preparation.
For a given magnitude, we run our method for all sample stars three times, and average the resulting three values for the Bayes factor of each star, to mitigate statistical deviations. The resulting averaged Bayes factor values for each star and magnitude are again averaged, giving us the central value $\overline{O}_{osc,noise}$ and uncertainty for a given magnitude. Using this value, we can then estimate up to which magnitude, the oscillations are still detectable.

\begin{figure}[!htb]
    \centering
    \includegraphics[width=\linewidth]{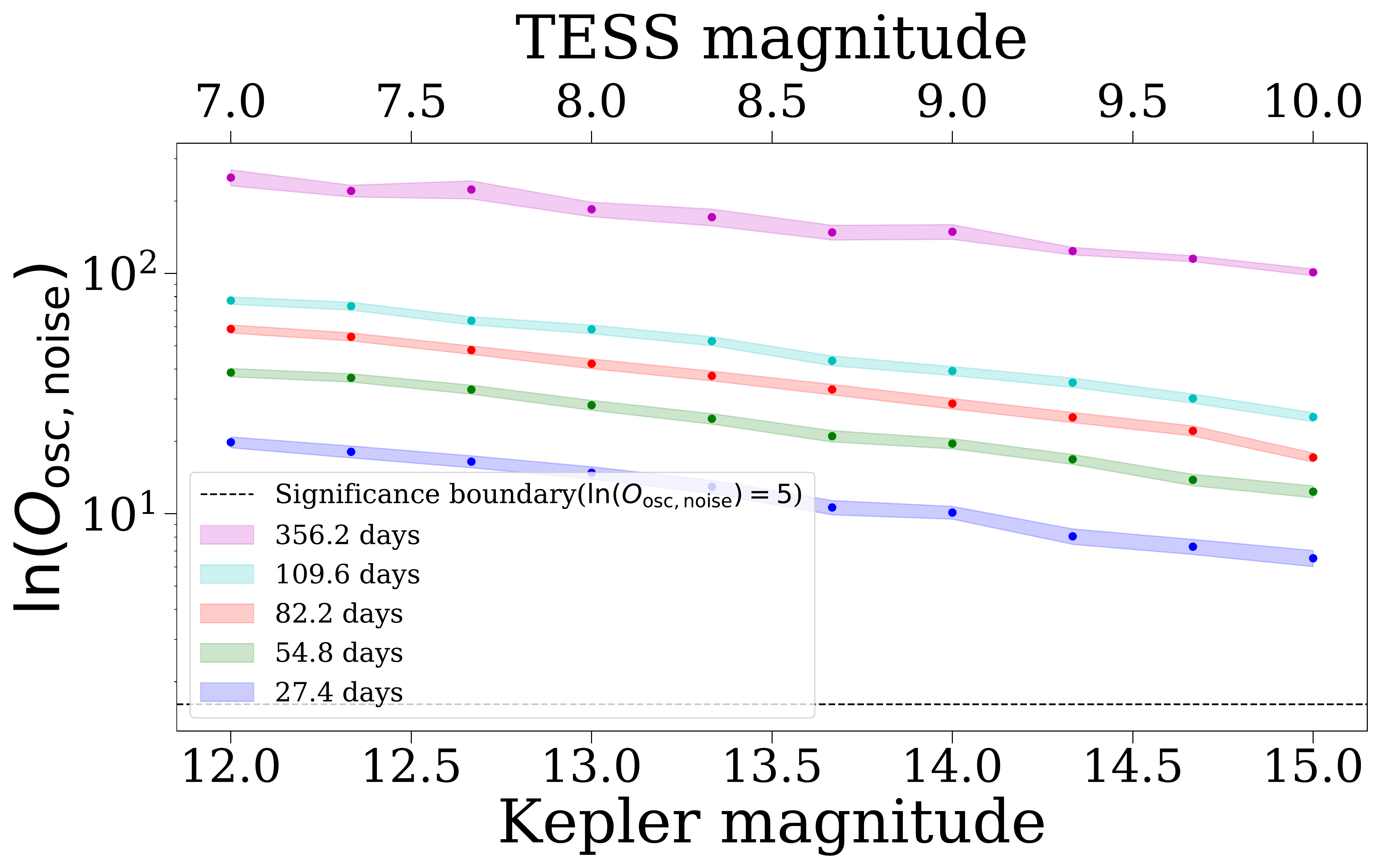}
    \caption{Relation between Bayes factor $O_\text{osc,noise}$ and \kepler\, magnitude as well as the converted TESS magnitude. Different observation lengths are color coded.}
    \label{fig:magnitude_analysis}
\end{figure}

The result is illustrated in Fig.~\ref{fig:magnitude_analysis}. As expected, we find a decrease in $\overline{O}_{osc,noise}$, proportional to the observation length of the data sets. The Bayes factor is highest for bright objects with long observation baselines, which is a natural result of having a better signal to noise and a higher frequency resolution in the PSD of the star than for fainter magnitudes and shorter observing lengths. Longer observation times give us a stronger shape of the power excess and an equal increase in the significance of the full background model. Despite of this, $O_{osc,noise}$ still exceeds the significance threshold of $\ln(O_{osc,noise})=5$ for all observation lengths and magnitudes up to $Kp=15$\,mag ($I_C=10\,\mbox{mag}$), making it in principle possible to detect solar-like oscillations in faint red giants up to $Kp=15$\,mag ($I_C=10\,\mbox{mag}$) even with an observation length of $27.4$\,d.


\begin{figure}[!htb]
    \centering
    \includegraphics[width=\linewidth]{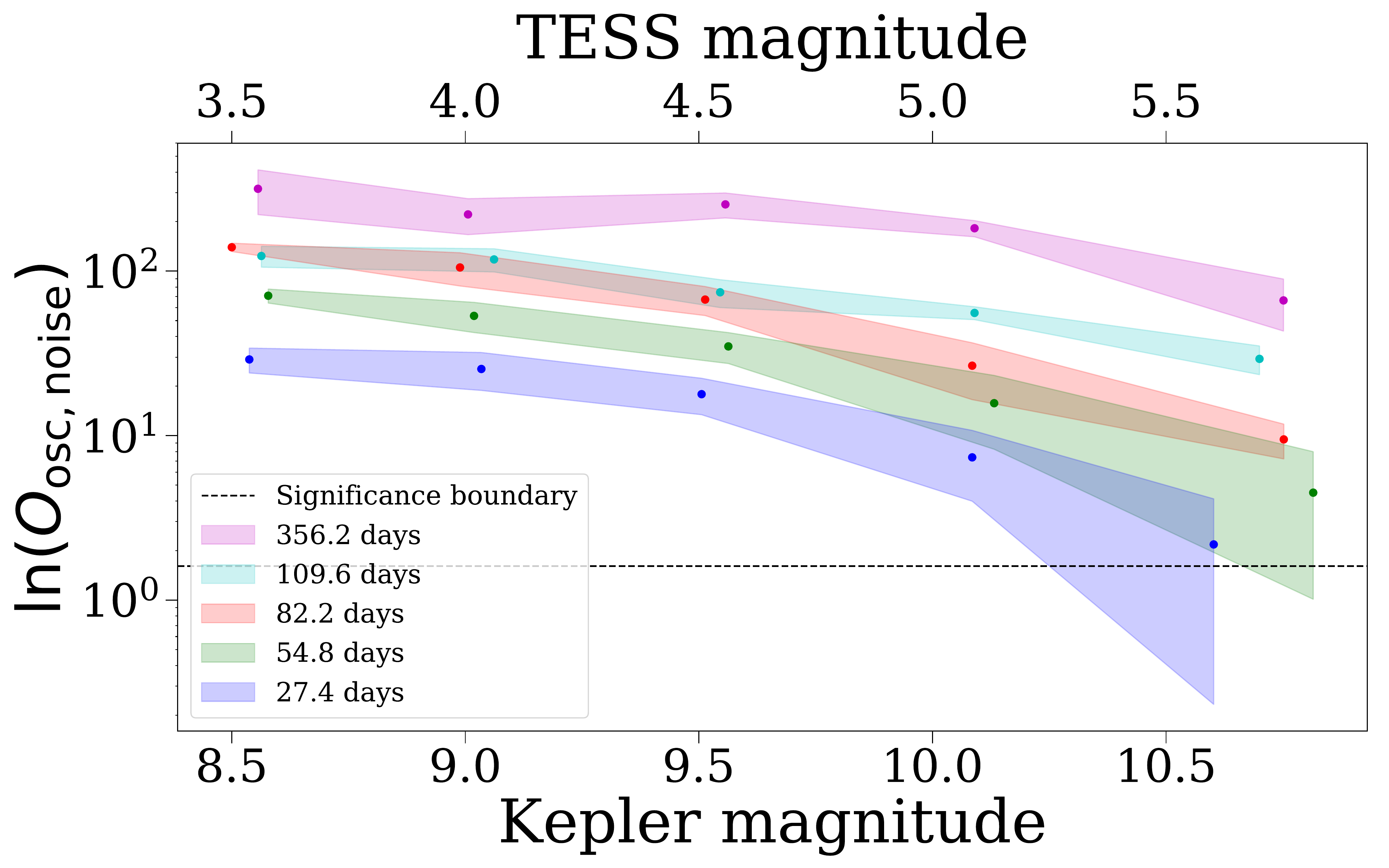}
    \caption{Relation between Bayes factor $O_\text{osc,noise}$ and \kepler\, magnitude for the main sequence sample.}
    \label{fig:legacy_bayes}
\end{figure}

A similar analysis was also done for the main sequence sample, except in this case we used the complete sample. This clearly shows a different picture than for red giants as illustrated in Fig.~\ref{fig:legacy_bayes}. The main sequence sample is much more sensitive to the brightness of the object, and already approaching the significance threshold for observation lengths $T_\text{obs}=\SI{54.8}{days}$ for fainter stars. This is due to the fact, that main-sequence solar like oscillators are much fainter, and following the luminosity-mass and amplitude relation \citep{Corsaro2013}, show a much lower oscillation amplitude. Therefore, it is intrinsically harder to find solar like oscillation in main-sequence stars than in red giants.

\section{Searching for pre-MS solar like oscillators}
\subsection{Pre-MS models}\label{sec:models}
We use version r-11701 of the software Modules for Experiments in Stellar Astrophyscis, \texttt{MESA}, \citep{Paxton2011, Paxton2013, Paxton2015, Paxton2018, Paxton2019} to construct a set of non-rotating stellar models with initial masses between $0.2$ and $5.15\,M_\odot$ and initial abundances $Z= 0.02$ and $Y = 0.28$. These stellar models are evolved from the Hayashi tracks until the zero-age main sequence (ZAMS). We choose to evolve the models without mass loss or accretion and with an Eddington-Grey atmosphere. The treatment of convection is via the mixing length theory with mixing length $\alpha_\text{MLT} = 1.8$ and convective premixing, introduced in \citet{Paxton2019}.

\begin{figure}
    \centering
    \includegraphics[width=\linewidth]{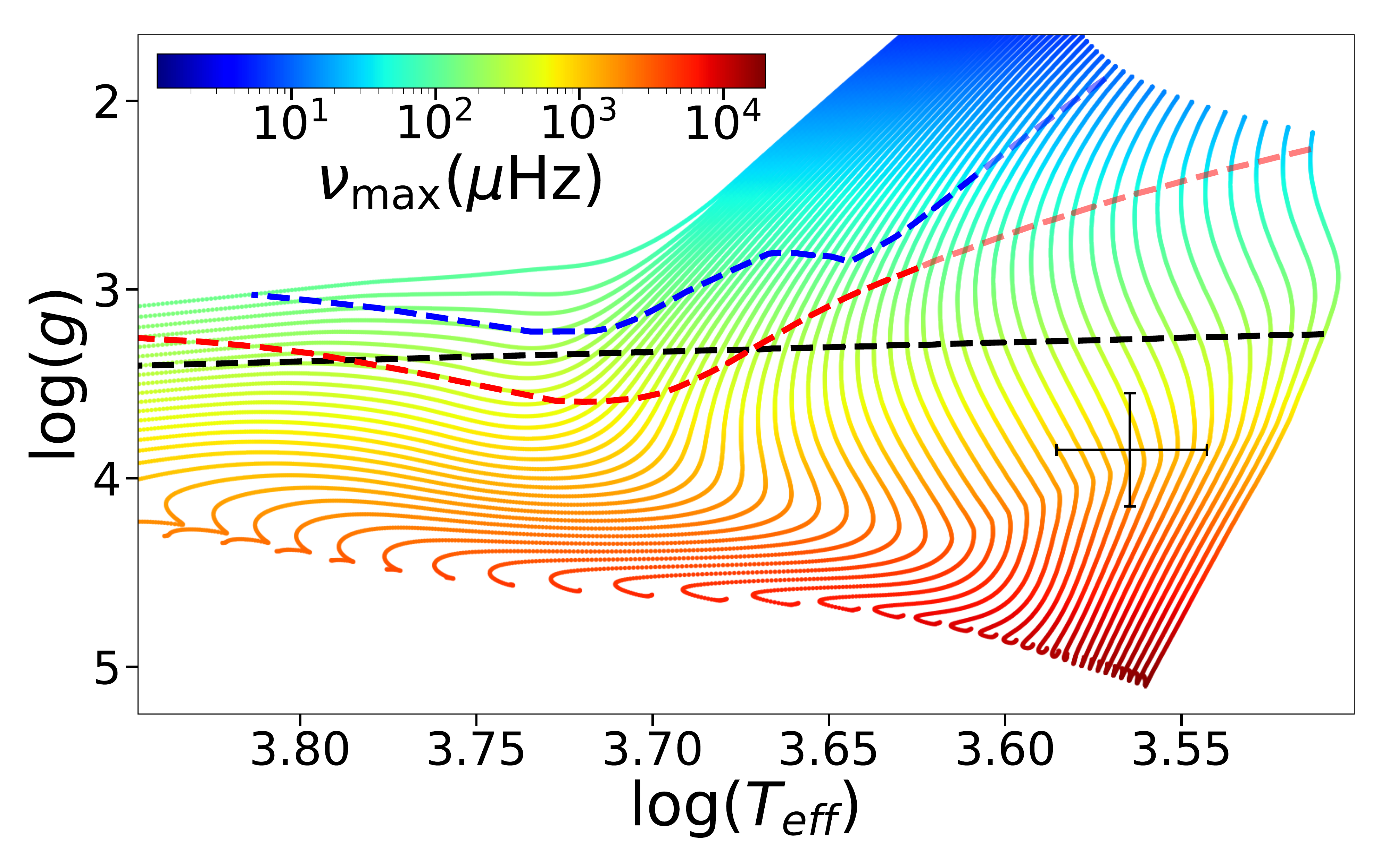}
    \caption{The evolutionary tracks in a $\log T_\mathrm{eff}$ - $\log{g}$ diagram. The color code describes $\numax$ obtained from the scaling relation. 
    The black line shows the location at which $\numax$ is equal to the Nyquist frequency of LC data . The red and blue lines show the locations of the birthline according to \cite{Palla1990} and \cite{Behrend2001}, respectively. The black crossed object shows the spectroscopic observations for EPIC 205375290.}
    \label{fig:premstracks}
\end{figure}

Using the scaling relation 
\begin{equation}\label{eq:numaxscalingrelation}
    \numax = \Bigg(\frac{g}{g_\odot}\Bigg)\Bigg(\frac{T_\text{eff}}{T_{\text{eff,}\odot}}\Bigg )^{-1/2}\nu_{\mathrm{max,}\odot}
\end{equation}
we can estimate $\numax$ for the computed stellar models and compare with the Nyquist frequency of LC data. The Nyquist frequency represents a natural upper limit to the detectable $\numax$ values. We therefore locate the position in the $\log T_\mathrm{eff}$ - $\log{g}$ diagram at which the estimated $\numax$ attains the Nyquist frequency of LC data. The results are shown in Fig.~\ref{fig:premstracks}, which also shows two versions of the location of the birthline \footnote{The blended parts of the birthlines are extrapolated to ensure that the birthlines cover all tracks.} according to the description of \cite{Palla1990} and \cite{Behrend2001}. 

The birthline describes the evolutionary track of an accreting protostar. After the star has accreted its final mass, it evolves along the evolutionary tracks shown in Fig.~\ref{fig:premstracks}. Consequently, it is very unlikely to observe stars that lie above the birthline. Therefore, LC data only allows the detection of solar like oscillations in pre-MS stars for the small portion of evolutionary tracks that lie below the birthline and above the black dashed line showing the Nyquist frequency in Figure \ref{fig:premstracks}.

\subsection{Pre-MS solar-like candidates}\label{sec:candidate_selection}
In a first step we apply the Everest pipeline by \cite{Luger2016,Luger2017} on our sample of 135 pre-MS stars to remove instrumental noise. The oscillation frequencies of optically visible pre-MS solar like oscillators are expected to be above $\SI{100}{\mu Hz}$ (see Fig.~\ref{fig:premstracks}). Thus, we can safely filter out the frequencies in the lower frequency ranges that are caused by interactions between the star and its surrounding disk as well as activity in the star. This is also necessary, as the amplitude of these low frequency variations are significantly higher than the expected amplitude of the oscillations. Not removing these variations would make it impossible to detect potential oscillations in our sample. We therefore apply a Savitzky-Golay (Savgol) filter \citep{Savitzky1964} and choose a third-order polynomial with a window size between 41 and 331, depending on the individual data set. 

\begin{figure*}[!htb]
    \centering
    \includegraphics[width=\textwidth]{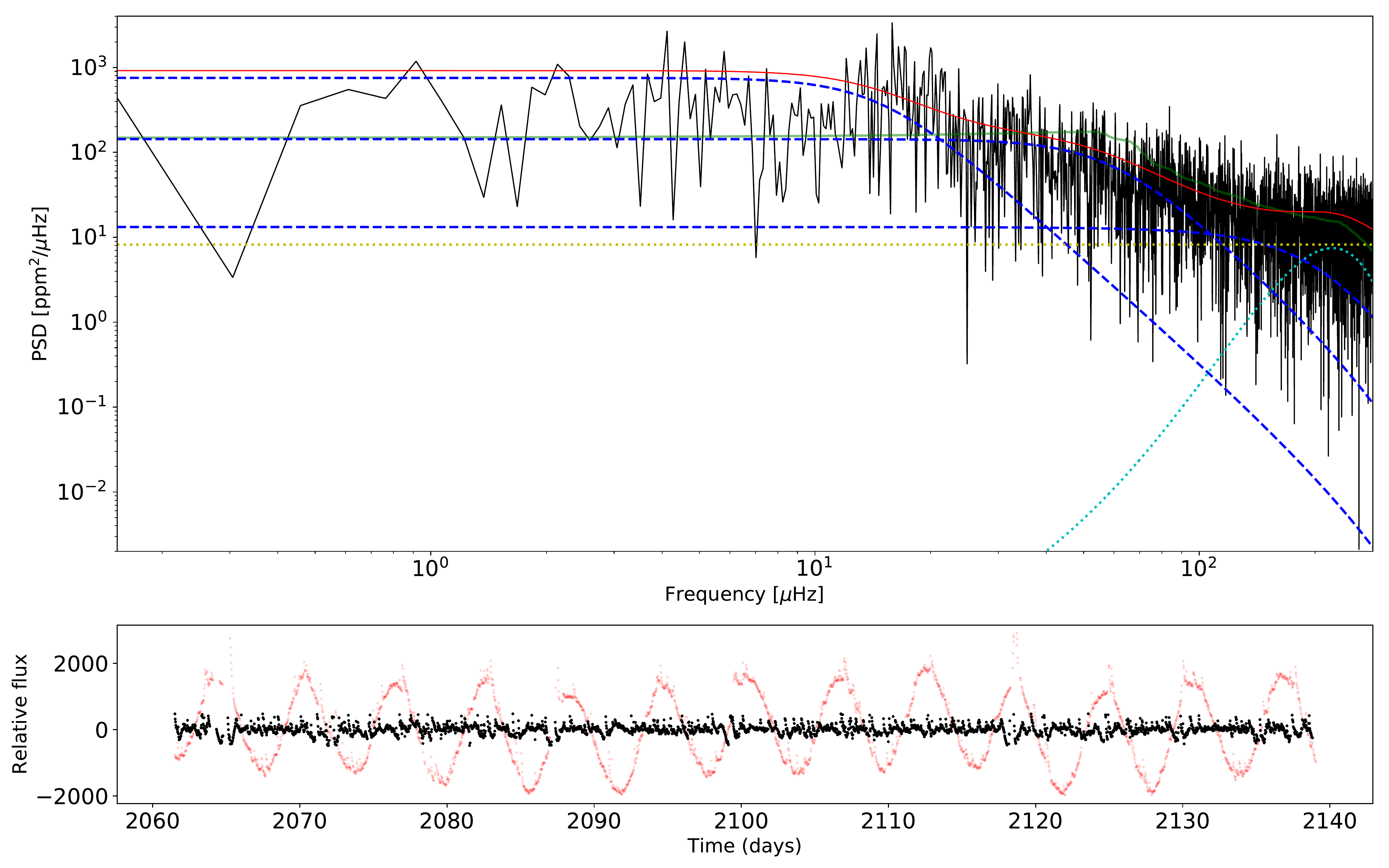}
    \caption{Upper panel: The resulting fit for EPIC 205375290. In dashed blue are the Harvey-like functions, the dotted cyan line shows the power excess, the yellow dotted line the white noise and the red solid line the full fit of the oscillation model. The green line represents a smoothed variant of the PSD. Lower panel: The light curve of EPIC 205375290. In black the filtered and reduced light curve, used for the analysis. In red the original light curve.}
    \label{fig:205375290}
\end{figure*}

In the next step, we use the FliPer method (see Section~\ref{sec:nu_max_determination}) and the scaling relation from equation \ref{eq:numaxscalingrelation} under the assumption that both relations are valid for pre-MS solar like oscillations, and estimate the stars' surface gravity to $\logg_\text{FliPer}$. We then compare $\logg_\text{FliPer}$ to $\logg_\text{EPIC}$ from the EPIC catalogue for every star. If we can find agreement between $\logg_\text{FliPer}$ and $\logg_\text{EPIC}$, we mark that star as a potential pre-MS solar-like candidate. This requirement is fullfilled by 13 stars which are listed in Table \ref{Tab:results}.

As expected, most of the candidates have an expected $\numax$ larger than the Nyquist frequency of $\SI{258.4}{\mu Hz}$ for LC K2 data. As no SC observations are available for these targets, we restrict our analysis to the narrow region between the Nyquist frequency and the birthline(s) as illustrated in Figure~\ref{fig:premstracks}. This leaves us with five objects for further inspection.


We apply \apollo\, to each of these five targets five times to reduce statistical fluctuations. The results for the five stars are listed in Table \ref{Tab:results}. The stars EPIC 204330922, 205152548 and 204222295 are rejected by \apollo\, and show a $\ln(O_{osc,noise})<0$, indicating that there is no significant indication of oscillation according to our model comparison approach. This either means that the oscillation signal is not strong enough for us to detect oscillations in these stars, or that these stars do not show solar-like oscillations.

For both EPIC 204447221 and 205375290 \apollo\, yields a Bayes factor $\ln(O_{osc,noise})$ that exceeds the significance threshold $\ln(O_{osc,noise})>5$. To exclude an induced signal through the treatment of the original light curve with the Savgol filter, we increase/decrease the window size by 10 and redo the analysis. In all cases the Bayes factor exceeds the significance threshold. We find disagreement between the final value of $\logg_\apollo$ and $\logg_\mathrm{EPIC}$, except for EPIC 205375290, which is why we further want to focus on this star. The resulting fit and light curve for EPIC 205375290 are shown in Fig.~\ref{fig:205375290}.

\begin{table*}
\caption{\apollo\, results for all LC candidates.}             
\label{Tab:results}      
\centering                                      
\begin{tabular}{c|c c c c c c}          
\hline\hline                      
EPIC ID&$\nu_\text{max,FliPer}$ &$\logg_\text{FliPer}$ &$\logg_\text{EPIC}$ &$\nu_\text{max,\apollo}$ &$\ln(O_{osc,noise})$&$\logg_\text{\apollo}$ \\ 
 & $[\mu$Hz] & $[$dex$]$ &  $[$dex$]$ & $[\mu$Hz] & & $[$dex$]$\\ 
\hline                                   
\begin{tabular}[c]{@{}l@{}}204330922\\205152548\\204222295\\204447221\\205375290\\205188906\\204854345\\203756781\\205359167\\205179845\\204121833\\204933717\\205078547\end{tabular} & 
\begin{tabular}[c]{@{}l@{}}105\\149\\150\\185\\219\\322\\554\\572\\983\\2212\\2295\\2295\\2560\end{tabular} & 
\begin{tabular}[c]{@{}l@{}}2.0\\2.2\\2.2\\2.3\\3.4\\2.5\\2.7\\2.8\\3.0\\3.3\\3.4\\3.4\\3.4\end{tabular} & 
\begin{tabular}[c]{@{}l@{}}2.3(9)\\2.4(4)\\2.4(6)\\2.1(5)\\3.2(1.5)\\3.2(7)\\2.5(7)\\2.6(4)\\2.7(3)\\4.3(1.1)\\4.1(9)\\4.1(9)\\4.1(1.4)\end{tabular} &
\begin{tabular}[c]{@{}l@{}}109(5)\\151(7)\\177(14)\\225(16)\\242(10)\\-\\-\\-\\-\\-\\-\\-\\-\end{tabular} & 
\begin{tabular}[c]{@{}l@{}}-20(16)\\-4.09(23)\\-77.26(23)\\6.55(22)\\9.07(25)\\-\\-\\-\\-\\-\\-\\-\\-\end{tabular} & 
\begin{tabular}[c]{@{}l@{}}2.94(10)\\3.09(10)\\3.16(11)\\3.26(11)\\3.30(10)\\-\\-\\-\\-\\-\\-\\-\\-\end{tabular} \\
\hline                                             
\end{tabular}
\tablefoot{The first three columns show the values for $\numax$ and $\logg$ from FliPer, as well as $\logg$ from the EPIC catalogue by \cite{Huber2016}. To compute $\nu_\text{max,\apollo}$, $\logg_\text{\apollo}$ and $O_{osc,noise}$ we take the mean over the 5 successive runs of the \apollo\, pipeline. The Bayes factor $O_{osc,noise}$ indicates a significant result, if it exceeds a value of $\ln(O_{osc,noise})>5$ (see Table \ref{Tab:Evidence})}
\end{table*}

\subsubsection{EPIC 205375290:  a possible candidate}

EPIC\,205375290 (2MASS\,J16111534-1757214) is a bona-fide member of the Upper-Scorpius association (RA$_{2000}$\,=\,16:11:15.34,   DE$_{2000}$\,=\,-17:57:21.42) and is listed in SIMBAD as a star of spectral type M1 \citep{Preibisch2001,Pecaut2016}. The object has a $V$ magnitude of 14.099 \citep{Pecaut2016} and a $J$ magnitude of 10.227 \citep{Cutri2003} illustrating a higher flux in the near-infrared.

The \apollo\ pipeline yields a $\nu_\text{max}$ of 242(10)\,$\mu$Hz and exceeds our significance threshold with $\ln(O_{osc,noise})=9.07(25)$. From the scaling relations, we calculate a surface gravity of $\logg=3.3(1)$.

To be able to place EPIC 205375290 in the HR-diagram and learn more about the star's properties, we first analyzed a low resolution (R=3200) optical spectrum obtained with the RSS spectrograph \citep[]{Burgh2003, Kobulnicky2003} mounted at the South African Large Telescope \citep[SALT;][]{salt2006}. We identify \ion{Li}{i} 6708\,\AA\ to be in absorption and found emission in the hydrogen Balmer lines and \ion{Ca}{ii} H\&K lines. These features are superimposed on a late-type spectrum with prominent \ion{TiO}{} absorption bands. Such a spectroscopic pattern is in agreement with the young evolutionary stage of EPIC 205375290. Comparing the depths and shapes of the \ion{TiO}{} bandheads with the template spectra from the MILES library \citep{Sanchez2006}, we confirm the spectral class of EPIC 205375290 to be dM1e, which is in agreement with the previous determination by \citet{Preibisch2001}.

\begin{figure}
    \centering
    \includegraphics[width=\linewidth]{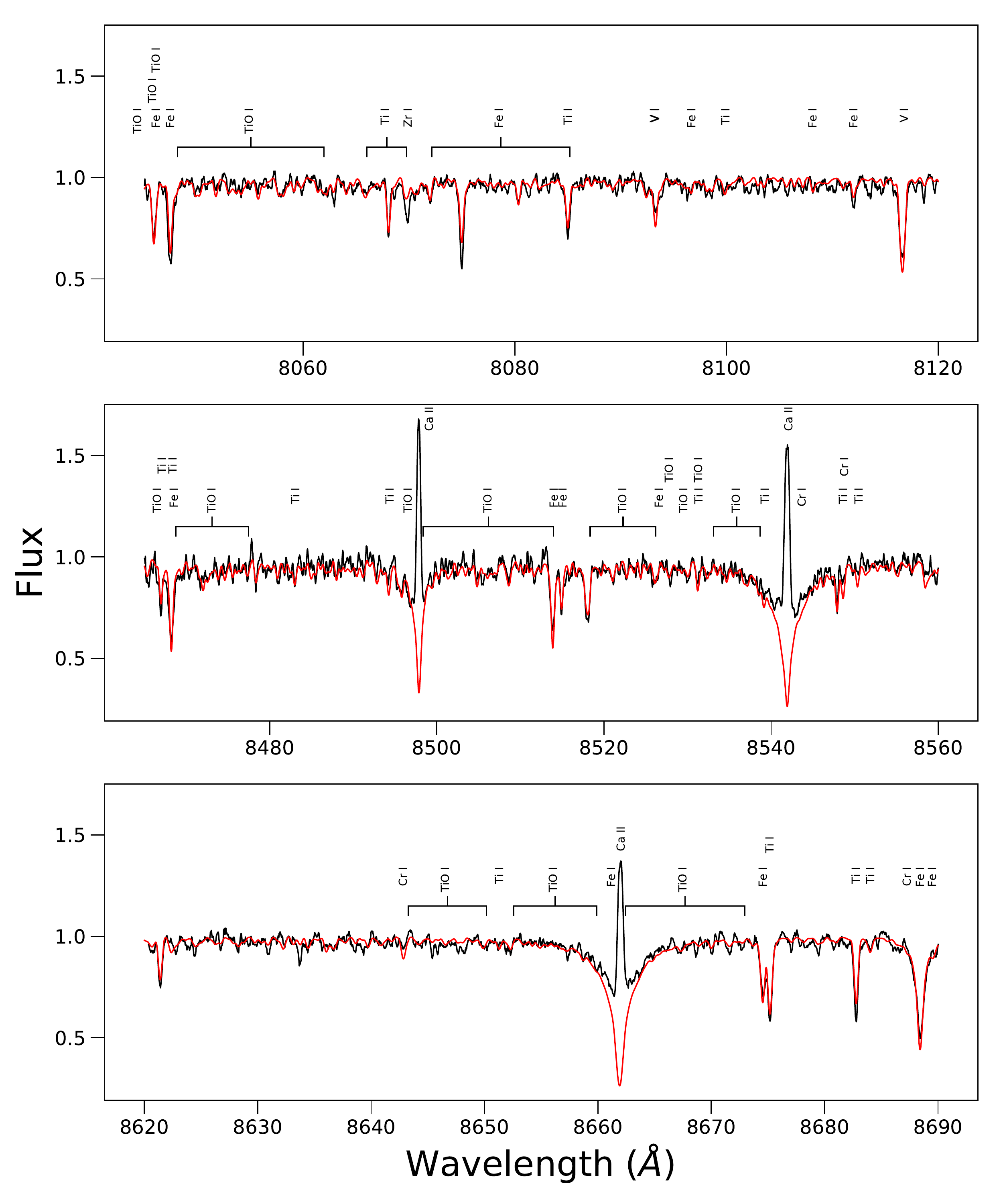}
    \caption{Regions from 8040 to 8120\,\AA\ (top panel), 8415 to 8555\,\AA\ (middle panel), and 8620 to 8695\,\AA\ (bottom panel) of the HIRES Keck spectrum of EPIC 205375290: the observed spectrum is shown in black and the calculated synthetic spectrum with the final adopted parameters of $T_\mathrm{eff}$, of 3670$\pm$180\,K and a log\,$g$ of 3.85$\pm$0.3 in red.}
    \label{fig:spectrum}
\end{figure}

We then continued our analysis with a high-resolution optical and near-infrared spectrum obtained by the HIRES (High Resolution Echelle Spectrometer) spectrograph at the Keck 1 telescope which is available in the Keck archive\footnote{https://www2.keck.hawaii.edu/koa/public/koa.php}. The observations were taken on June 16 2006 with the C5 slit which has a width of 1.148" projected onto the sky resulting in a nominal spectral resolution of R=38000. The spectrum covers the range between 4800 and 9200\,\AA\, with significant inter-order gaps in the red. It has a signal-to-noise ratio of about 35 per pixel (or 80 per resolution element). In addition to the general spectral patterns identified with the low-resolution spectroscopy, the HIRES spectrum revealed the double-peaked profile of $\mathrm{H}\alpha$ emission, a complex structure in the cores of the resonance \ion{Na}{i} D lines, emission in the \ion{He}{i} line at 5876\,\AA\, and emission in the \ion{Ca}{ii} infrared lines. All these spectroscopic features clearly indicate that EPIC 205375290 is an accreting T Tauri star \citep{Edwards1994}.

The photospheric spectra of T Tauri stars are known to be subject to the veiling effect \citep[e.g.][]{Basri1990,Dodin2012} which can prevent the accurate determination of the atmospheric parameters using the absorption lines. The equivalent width of the $\mathrm{H}\alpha$ emission in EPIC 205375290 is 3\,\AA\ and its width at 10\% intensity is about $\SI{180}{\km/\s}$ derived from the HIRES spectrum observed in 2006. According to the criterion from \cite{White2003}, this leads us to conclude that the star is a weak accretor and has small veiling, if any. 
However, the slightly redshifted \ion{He}{i} 5876\,\AA\, emission (i.e. about $\SI{+5}{\km/\s}$ in respect to the stellar rest velocity) indicates the presence of an accretion shock in the system and does not allow us to completely ignore the possible contribution of the non-photospheric continuum or line emission to the observed spectrum of EPIC 205375290. To check this, we fitted the \ion{He}{i} 5876\,\AA\, line and the accretion-related emission cores in the infrared \ion{Ca}{ii} triplet (which has the same redshift as the He line) with the pre-computed spectra from the grid of non-LTE models of accretion shocks in the atmospheres of young stars \citep[A. Dodin pers.comm.]{Dodin2015}. The best-fit model which adequately reproduces the observed widths and intensities of accretion-related tracers has an accretion rate of the order $\SI{e-11}{M\textsubscript{\(\odot\)}/\year}$ and shows no significant veiling of the photospheric lines in the red orders of the HIRES spectrum. Thus, the regions around the infrared \ion{Ca}{ii} triplet were chosen for further analysis.

We also used a high-resolution (R=47000) spectrum of EPIC 205375290 that was obtained with the red arm of the FLAMES-UVES spectrograph mounted at the ESO VLT UT2 telescope.
The observations were carried out on 1 June 2016 under programme 097.C-0040(A) and cover the wavelength range from 4800 to 6700\,\AA\ with an $\sim$150\,\AA\ wide central gap between the detectors. A comparison to the HIRES spectrum revealed the variability of the $\mathrm{H}\alpha$ emission: its equivalent width increased up to 4.1\,\AA\ in the FLAMES spectrum. Most probably such a variability reflects the variations in the accretion rate.
A detailed study of the activity of EPIC 205375290 will be subject of a future work.


For the determination of the star's fundamental parameters, we used the three HIRES red orders from 8040 -- 8120\,\AA, 8415 -- 8555\,\AA, and 8620 -- 8695\,\AA\, which contain several atomic lines including the important \ion{Ca}{ii} infrared triplet at 8498\,\AA, 8542\,\AA, and 8662\,\AA, because this region is almost free from veiling and contamination by the strong molecular bands. We avoid to use the \ion{TiO}{} bands despite their extreme temperature sensitivity because the current state of atomic data for these transitions complicates their use in quantitative analysis \citep{Valenti1998} . We determined the atmospheric parameters of EPIC 205375290 based on spectral synthesis of the atomic absorption lines using the SME software \citep{Valenti1996,Piskunov2017}. Our analysis yields an effective temperature, $T_\mathrm{eff}$, of 3670$\pm$180\,K, a log\,$g$ of 3.85$\pm$0.3, a $v$sin$i$ of 8 $\pm$ 1 km\,s$^{-1}$, and about solar metallicity. This allows us to mark EPIC 205375290's position in the Kiel diagram shown in Figure \ref{fig:premstracks}.

The accuracy of our determination of EPIC 205375290's fundamental parameters given in Table \ref{Tab:epic_fundpars} is limited by several factors: (i) The spectrum only has a modest signal-to-noise ratio of only $\sim$35. (ii) Numerous weak \ion{TiO}{} lines are present in the spectrum which are not well described by the current atomic data, hence they introduce additional ``noise''. (iii) The most important gravity indicators in M-type stars -- the regions of the \ion{K}{i} and red \ion{Na}{i} doublets at 7664\,\AA\ and 7699\,\AA\, and 8183\,\AA\ and 8194\,\AA\, respectively -- fell into inter-order gaps in the HIRES spectrum we used. Therefore, we used the wings of the \ion{Ca}{ii} lines as constraints for log\,$g$. (iv) Some of the spectral lines show a slight blueward asymmetry which could be an effect of the circumstellar environment or of a potential, yet unresolved binarity. Consequently, the derived fundamental parameters can only be used as a first estimate with relatively high errors. For a more detailed and more accurate analysis, a high-resolution, high signal-to-noise ratio spectrum would be required and will hopefully be obtained in the future.

It is obvious that our spectroscopically determined values for $T_\mathrm{eff}$ and log\,$g$ place EPIC 205375290 below the line where $\numax$ is equal to the Nyquist frequency of LC data in Figure \ref{fig:premstracks}. Currently, we can only speculate about the origin of this discrepancy. 
One explanation is related to the above mentioned high uncertainties of the spectroscopically determined fundamental parameters; in particular the log\,$g$ value is likely overestimated. Another reason for this discrepancy might come from the assumption that the scaling relation for stars that are on the main-sequence or later is fully applicable to pre-MS stars in an inverse form, which might not be the case. 
Additional work is required to investigate this further, but this is out of the scope of the present analysis.


\begin{table}
\caption{Fundamental parameters for EPIC 205375290 from high-resolution spectroscopy}             
\label{Tab:epic_fundpars}      
\centering                                      
\begin{tabular}{lrl}          
\hline
Quantity & Value & Unit \\
\hline                      
$T_\mathrm{eff}$ & 3670 $\pm180$ & K \\
log\,$g$ & 3.85 $ \pm 0.3$ & dex\\
$v$sin$i$ & 8.0 $ \pm 1.0$ & km\,s$^{-1}$ \\
$[$Fe/H$]$ & +0.02 $ \pm 0.12 $ & dex \\
$v_{\rm micro}$ & 1.7 $ \pm 0.5$ & km\,s$^{-1}$ \\
\hline                                   
\end{tabular}
\end{table}

\section{Conclusions}
We have developed \apollo, a software package based on \diamonds\ that uses a new Bayesian approach for the detection of low signal-to-noise solar-like oscillations in the presence of a high background level. High background levels can, for example, originate from highly active stars, in particular in their earliest stages of evolution prior to the onset of core-hydrogen burning. For the search of solar-like oscillations in pre-MS stars, a reliable treatment of such high background levels is a crucial prerequisite.

In addition to the photometric time series obtained from space, \apollo\ only requires the stars' effective temperatures and magnitudes as input for the determination of the frequency of maximum power, $\numax$, and the large frequency separation, $\Delta \nu$. For any Bayesian method, good priors have to be selected. For this, \apollo\ uses dedicated estimation algorithms which compute prior distributions for the ten parameters $\sigma_b$, $a_1$, $a_2$, $a_3$, $b_1$, $b_2$, $b_3$, $H_{\rm osc}$, $\numax$, and $\sigma_{\rm env}$. 

We tested our new method on 1071 confirmed solar like oscillators observed by the \kepler\ satellite in long cadence; the majority of the objects have data from the full four years of \kepler\ observations. With these data, we calibrated \apollo, and tested the reliability of our method. We then subsequently reduced the observation lengths to the time bases obtained with the TESS mission, which are 356.2\,d, 109.6\,d, 82.2\,d, 54.8\,d, and 27.4\,d. As the observational time bases get shorter, the distributions of the $\numax$ and $\Delta \nu$ values get broader. But overall, we find that our method is robust even when shortening the observation length dramatically. Consequently, \apollo\ can be applied to data observed by various current and future missions (e.g. TESS, \kepler\, K2 and in the future PLATO), and is designed such that it can process large amounts of observations quickly and efficiently.

In a final step, we applied \apollo\ to a sample of 135 candidate stars in the pre-MS stage observed in the K2 Campaign 2 on Upper Scorpius in long cadence. Most of the candidates have an expected $\numax$ larger than the Nyquist frequency for K2 LC data, and no SC observations are available for these stars. Although this complicates our search, we identified a possible candidate for pre-MS solar like oscillations, EPIC 205375290. It shows a power excess around a $\numax$ of 242 $\pm$ 10\,$\mu$Hz, which we found statistically signifcant acording to our Bayesian model comparison. For the verification of EPIC 205375290 as the first pre-MS solar like oscillator additional work is required in the future: on the observational side longer photometric time series at shorter cadence and higher signal-to-noise ratio spectroscopic data are needed; on the theoretical side, it has to be tested if the scaling relation used, which are usually applied to main-sequence stars and later, can be applied in its inverse form for pre-MS solar like oscillators.

\begin{acknowledgements}
E.C. is funded by the European Union’s Horizon 2020 research and innovation program under the Marie Sklodowska-Curie grant agreement No. 664931.

A.\,K. acknowledges support from the National Research Foundation (NRF) of South Africa.

I.P. was supported by budgetary funding of the Basic Research programme II.16.

Some spectral data for this work were obtained with the Southern African Large Telescope (SALT)
under program \mbox{2019-1-MLT-002}.
\end{acknowledgements}
\bibliographystyle{aa}
\small{\bibliography{references}}

\begin{thebibliography}{92}
\expandafter\ifx\csname natexlab\endcsname\relax\def\natexlab#1{#1}\fi

\bibitem[{{Alencar} {et~al.}(2010){Alencar}, {Teixeira}, {Guimar{\~a}es},
  {McGinnis}, {Gameiro}, {Bouvier}, {Aigrain}, {Flaccomio}, \&
  {Favata}}]{alencar2010}
{Alencar}, S.~H.~P., {Teixeira}, P.~S., {Guimar{\~a}es}, M.~M., {et~al.} 2010,
  \aap, 519, A88

\bibitem[{{Anderson} {et~al.}(1990){Anderson}, {Duvall}, \&
  {Jefferies}}]{Anderson1990}
{Anderson}, E.~R., {Duvall}, Thomas~L., J., \& {Jefferies}, S.~M. 1990, \apj,
  364, 699

\bibitem[{{Antoci} {et~al.}(2011){Antoci}, {Handler}, {Campante}, {Thygesen},
  {Moya}, {Kallinger}, {Stello}, {Grigahc{\`e}ne}, {Kjeldsen}, {Bedding},
  {L{\"u}ftinger}, {Christensen-Dalsgaard}, {Catanzaro}, {Frasca}, {De Cat},
  {Uytterhoeven}, {Bruntt}, {Houdek}, {Kurtz}, {Lenz}, {Kaiser}, {van Cleve},
  {Allen}, \& {Clarke}}]{antoci2011}
{Antoci}, V., {Handler}, G., {Campante}, T.~L., {et~al.} 2011, \nat, 477, 570

\bibitem[{{Appourchaux}(2014)}]{Appourchaux2014}
{Appourchaux}, T. 2014, {A crash course on data analysis in asteroseismology}
  (Cambridge University Press), 123

\bibitem[{{Appourchaux} {et~al.}(2012){Appourchaux}, {Chaplin}, {Garc{\'\i}a},
  {Gruberbauer}, {Verner}, {Antia}, {Benomar}, {Campante}, {Davies},
  {Deheuvels}, {Handberg}, {Hekker}, {Howe}, {R{\'e}gulo}, {Salabert},
  {Bedding}, {White}, {Ballot}, {Mathur}, {Silva Aguirre}, {Elsworth}, {Basu},
  {Gilliland }, {Christensen-Dalsgaard}, {Kjeldsen}, {Uddin}, {Stumpe}, \&
  {Barclay}}]{Appourchaux2012}
{Appourchaux}, T., {Chaplin}, W.~J., {Garc{\'\i}a}, R.~A., {et~al.} 2012, \aap,
  543, A54

\bibitem[{{Astropy Collaboration} {et~al.}(2013){Astropy Collaboration},
  {Robitaille}, {Tollerud}, {Greenfield}, {Droettboom}, {Bray}, {Aldcroft},
  {Davis}, {Ginsburg}, {Price-Whelan}, {Kerzendorf}, {Conley}, {Crighton},
  {Barbary}, {Muna}, {Ferguson}, {Grollier}, {Parikh}, {Nair}, {Unther},
  {Deil}, {Woillez}, {Conseil}, {Kramer}, {Turner}, {Singer}, {Fox}, {Weaver},
  {Zabalza}, {Edwards}, {Azalee Bostroem}, {Burke}, {Casey}, {Crawford},
  {Dencheva}, {Ely}, {Jenness}, {Labrie}, {Lim}, {Pierfederici}, {Pontzen},
  {Ptak}, {Refsdal}, {Servillat}, \& {Streicher}}]{astropy:2013}
{Astropy Collaboration}, {Robitaille}, T.~P., {Tollerud}, E.~J., {et~al.} 2013,
  \aap, 558, A33

\bibitem[{{Auvergne} {et~al.}(2009){Auvergne}, {Bodin}, {Boisnard}, {Buey},
  {Chaintreuil}, {Epstein}, {Jouret}, {Lam-Trong}, {Levacher}, {Magnan},
  {Perez}, {Plasson}, {Plesseria}, {Peter}, {Steller}, {Tiph{\`e}ne}, {Baglin},
  {Agogu{\'e}}, {Appourchaux}, {Barbet}, {Beaufort}, {Bellenger}, {Berlin},
  {Bernardi}, {Blouin}, {Boumier}, {Bonneau}, {Briet}, {Butler}, {Cautain},
  {Chiavassa}, {Costes}, {Cuvilho}, {Cunha-Parro}, {de Oliveira Fialho},
  {Decaudin}, {Defise}, {Djalal}, {Docclo}, {Drummond}, {Dupuis}, {Exil},
  {Faur{\'e}}, {Gaboriaud}, {Gamet}, {Gavalda}, {Grolleau}, {Gueguen},
  {Guivarc'h}, {Guterman}, {Hasiba}, {Huntzinger}, {Hustaix}, {Imbert},
  {Jeanville}, {Johlander}, {Jorda}, {Journoud}, {Karioty}, {Kerjean},
  {Lafond}, {Lapeyrere}, {Landiech}, {Larqu{\'e}}, {Laudet}, {Le Merrer},
  {Leporati}, {Leruyet}, {Levieuge}, {Llebaria}, {Martin}, {Mazy}, {Mesnager},
  {Michel}, {Moalic}, {Monjoin}, {Naudet}, {Neukirchner}, {Nguyen-Kim},
  {Ollivier}, {Orcesi}, {Ottacher}, {Oulali}, {Parisot}, {Perruchot},
  {Piacentino}, {Pinheiro da Silva}, {Platzer}, {Pontet}, {Pradines},
  {Quentin}, {Rohbeck}, {Rolland}, {Rollenhagen}, {Romagnan}, {Russ}, {Samadi},
  {Schmidt}, {Schwartz}, {Sebbag}, {Smit}, {Sunter}, {Tello}, {Toulouse},
  {Ulmer}, {Vandermarcq}, {Vergnault}, {Wallner}, {Waultier}, \&
  {Zanatta}}]{auvergne09}
{Auvergne}, M., {Bodin}, P., {Boisnard}, L., {et~al.} 2009, \aap, 506, 411

\bibitem[{{Basri} \& {Batalha}(1990)}]{Basri1990}
{Basri}, G. \& {Batalha}, C. 1990, \apj, 363, 654

\bibitem[{{Bastien} {et~al.}(2013){Bastien}, {Stassun}, {Basri}, \&
  {Pepper}}]{Bastien2013}
{Bastien}, F.~A., {Stassun}, K.~G., {Basri}, G., \& {Pepper}, J. 2013, \nat,
  500, 427

\bibitem[{{Behrend} \& {Maeder}(2001)}]{Behrend2001}
{Behrend}, R. \& {Maeder}, A. 2001, \aap, 373, 190

\bibitem[{{Bell} {et~al.}(2019){Bell}, {Hekker}, \& {Kuszlewicz}}]{Bell2019}
{Bell}, K.~J., {Hekker}, S., \& {Kuszlewicz}, J.~S. 2019, \mnras, 482, 616

\bibitem[{{Bonanno} {et~al.}(2014){Bonanno}, {Corsaro}, \&
  {Karoff}}]{Bonanno2014}
{Bonanno}, A., {Corsaro}, E., \& {Karoff}, C. 2014, \aap, 571, A35

\bibitem[{Borucki {et~al.}(2010)Borucki, Koch, Basri, Batalha, Brown, Caldwell,
  Caldwell, Christensen-Dalsgaard, Cochran, DeVore, Dunham, Dupree, Gautier,
  Geary, Gilliland, Gould, Howell, Jenkins, Kjeldsen, Kondo, Latham, Lissauer,
  Marcy, Meibom, Monet, Morrison, Sasselov, \& Tarter}]{Borucki2010}
Borucki, W.~J., Koch, D., Basri, G., {et~al.} 2010, in American Astronomical
  Society Meeting Abstracts, Vol. 215, American Astronomical Society Meeting
  Abstracts {\#}215, 101.01

\bibitem[{Brown {et~al.}(2011)Brown, Latham, Everett, \& Esquerdo}]{Brown2011}
Brown, T.~M., Latham, D.~W., Everett, M.~E., \& Esquerdo, G.~A. 2011,
  Astronomical Journal, 142

\bibitem[{{Buckley} {et~al.}(2006){Buckley}, {Swart}, \& {Meiring}}]{salt2006}
{Buckley}, D. A.~H., {Swart}, G.~P., \& {Meiring}, J.~G. 2006, in Society of
  Photo-Optical Instrumentation Engineers (SPIE) Conference Series, Vol. 6267,
  Society of Photo-Optical Instrumentation Engineers (SPIE) Conference Series,
  62670Z

\bibitem[{{Bugnet} {et~al.}(2018){Bugnet}, {Garc{\'\i}a}, {Davies}, {Mathur},
  {Corsaro}, {Hall}, \& {Rendle}}]{Bugnet2018}
{Bugnet}, L., {Garc{\'\i}a}, R.~A., {Davies}, G.~R., {et~al.} 2018, \aap, 620,
  A38

\bibitem[{{Bugnet} {et~al.}(2019){Bugnet}, {Garc{\'\i}a}, {Mathur}, {Davies},
  {Hall}, {Lund}, \& {Rendle}}]{Bugnet2019}
{Bugnet}, L., {Garc{\'\i}a}, R.~A., {Mathur}, S., {et~al.} 2019, \aap, 624, A79

\bibitem[{Burgh {et~al.}(2003)Burgh, Nordsieck, Kobulnicky, Williams,
  O'Donoghue, Smith, \& Percival}]{Burgh2003}
Burgh, E.~B., Nordsieck, K.~H., Kobulnicky, H.~A., {et~al.} 2003, in Instrument
  Design and Performance for Optical/Infrared Ground-based Telescopes, ed.
  M.~Iye \& A.~F.~M. Moorwood, Vol. 4841, International Society for Optics and
  Photonics (SPIE), 1463 -- 1471

\bibitem[{{Campante} {et~al.}(2016){Campante}, {Schofield}, {Kuszlewicz},
  {Bouma}, {Chaplin}, {Huber}, {Christensen-Dalsgaard}, {Kjeldsen}, {Bossini},
  {North}, {Appourchaux}, {Latham}, {Pepper}, {Ricker}, {Stassun},
  {Vanderspek}, \& {Winn}}]{Campante2016}
{Campante}, T.~L., {Schofield}, M., {Kuszlewicz}, J.~S., {et~al.} 2016, \apj,
  830, 138

\bibitem[{{Chaplin} {et~al.}(2011){Chaplin}, {Bedding}, {Bonanno}, {Broomhall},
  {Garc{\'\i}a}, {Hekker}, {Huber}, {Verner}, {Basu}, {Elsworth}, {Houdek},
  {Mathur}, {Mosser}, {New}, {Stevens}, {Appourchaux}, {Karoff}, {Metcalfe},
  {Molenda-{\.Z}akowicz}, {Monteiro}, {Thompson}, {Christensen-Dalsgaard},
  {Gilliland}, {Kawaler}, {Kjeldsen}, {Ballot}, {Benomar}, {Corsaro},
  {Campante}, {Gaulme}, {Hale}, {Handberg}, {Jarvis}, {R{\'e}gulo}, {Roxburgh},
  {Salabert}, {Stello}, {Mullally}, {Li}, \& {Wohler}}]{Chaplin2011}
{Chaplin}, W.~J., {Bedding}, T.~R., {Bonanno}, A., {et~al.} 2011, \apjl, 732,
  L5

\bibitem[{{Chaplin} \& {Miglio}(2013)}]{Chaplin2013}
{Chaplin}, W.~J. \& {Miglio}, A. 2013, \araa, 51, 353

\bibitem[{{Cody} {et~al.}(2014){Cody}, {Stauffer}, {Baglin}, {Micela},
  {Rebull}, {Flaccomio}, {Morales-Calder{\'o}n}, {Aigrain}, {Bouvier},
  {Hillenbrand}, {Gutermuth}, {Song}, {Turner}, {Alencar}, {Zwintz},
  {Plavchan}, {Carpenter}, {Findeisen}, {Carey}, {Terebey}, {Hartmann},
  {Calvet}, {Teixeira}, {Vrba}, {Wolk}, {Covey}, {Poppenhaeger}, {G{\"u}nther},
  {Forbrich}, {Whitney}, {Affer}, {Herbst}, {Hora}, {Barrado}, {Holtzman},
  {Marchis}, {Wood}, {Medeiros Guimar{\~a}es}, {Lillo Box}, {Gillen},
  {McQuillan}, {Espaillat}, {Allen}, {D'Alessio}, \& {Favata}}]{cody2014}
{Cody}, A.~M., {Stauffer}, J., {Baglin}, A., {et~al.} 2014, \aj, 147, 82

\bibitem[{{Colman} {et~al.}(2017){Colman}, {Huber}, {Bedding}, {Kuszlewicz},
  {Yu}, {Beck}, {Elsworth}, {Garc{\'\i}a}, {Kawaler}, {Mathur}, {Stello}, \&
  {White}}]{Colman2017}
{Colman}, I.~L., {Huber}, D., {Bedding}, T.~R., {et~al.} 2017, \mnras, 469,
  3802

\bibitem[{{Corsaro} \& {De Ridder}(2014)}]{corsaro2014}
{Corsaro}, E. \& {De Ridder}, J. 2014, \aap, 571, A71

\bibitem[{{Corsaro} \& {De Ridder}(2015)}]{Corsaro2015}
{Corsaro}, E. \& {De Ridder}, J. 2015, in European Physical Journal Web of
  Conferences, Vol. 101, European Physical Journal Web of Conferences, 06019

\bibitem[{{Corsaro} {et~al.}(2013){Corsaro}, {Fr{\"o}hlich}, {Bonanno},
  {Huber}, {Bedding}, {Benomar}, {De Ridder}, \& {Stello}}]{Corsaro2013}
{Corsaro}, E., {Fr{\"o}hlich}, H.~E., {Bonanno}, A., {et~al.} 2013, \mnras,
  430, 2313

\bibitem[{{Corsaro} {et~al.}(2017){Corsaro}, {Mathur}, {Garc{\'\i}a}, {Gaulme},
  {Pinsonneault}, {Stassun}, {Stello}, {Tayar}, {Trampedach}, {Jiang},
  {Nitschelm}, \& {Salabert}}]{Corsaro2017}
{Corsaro}, E., {Mathur}, S., {Garc{\'\i}a}, R.~A., {et~al.} 2017, \aap, 605, A3

\bibitem[{{Cutri} {et~al.}(2003){Cutri}, {Skrutskie}, {van Dyk}, {Beichman},
  {Carpenter}, {Chester}, {Cambresy}, {Evans}, {Fowler}, {Gizis}, {Howard},
  {Huchra}, {Jarrett}, {Kopan}, {Kirkpatrick}, {Light}, {Marsh}, {McCallon},
  {Schneider}, {Stiening}, {Sykes}, {Weinberg}, {Wheaton}, {Wheelock}, \&
  {Zacarias}}]{Cutri2003}
{Cutri}, R.~M., {Skrutskie}, M.~F., {van Dyk}, S., {et~al.} 2003, VizieR Online
  Data Catalog, II/246

\bibitem[{{De Ridder} {et~al.}(2009){De Ridder}, {Barban}, {Baudin}, {Carrier},
  {Hatzes}, {Hekker}, {Kallinger}, {Weiss}, {Baglin}, {Auvergne}, {Samadi},
  {Barge}, \& {Deleuil}}]{deridder2009}
{De Ridder}, J., {Barban}, C., {Baudin}, F., {et~al.} 2009, \nat, 459, 398

\bibitem[{{Dodin}(2015)}]{Dodin2015}
{Dodin}, A.~V. 2015, Astronomy Letters, 41, 196

\bibitem[{{Dodin} \& {Lamzin}(2012)}]{Dodin2012}
{Dodin}, A.~V. \& {Lamzin}, S.~A. 2012, Astronomy Letters, 38, 649

\bibitem[{{Duvall} {et~al.}(1986){Duvall}, {Harvey}, \&
  {Pomerantz}}]{Duvall1986}
{Duvall}, T.~L., {Harvey}, J.~W., \& {Pomerantz}, M.~A. 1986, Antarctic Journal
  of the U.S, 21, 280

\bibitem[{{Edwards} {et~al.}(1994){Edwards}, {Hartigan}, {Ghandour}, \&
  {Andrulis}}]{Edwards1994}
{Edwards}, S., {Hartigan}, P., {Ghandour}, L., \& {Andrulis}, C. 1994, \aj,
  108, 1056

\bibitem[{{Feroz} \& {Hobson}(2008)}]{Feroz2008}
{Feroz}, F. \& {Hobson}, M.~P. 2008, \mnras, 384, 449

\bibitem[{{Fr{\"o}hlich} {et~al.}(2012){Fr{\"o}hlich}, {Frasca}, {Catanzaro},
  {Bonanno}, {Corsaro}, {Molenda-{\.Z}akowicz}, {Klutsch}, \&
  {Montes}}]{Frohlich2012}
{Fr{\"o}hlich}, H.~E., {Frasca}, A., {Catanzaro}, G., {et~al.} 2012, \aap, 543,
  A146

\bibitem[{{Garc{\'\i}a} \& {Ballot}(2019)}]{Garcia2019}
{Garc{\'\i}a}, R.~A. \& {Ballot}, J. 2019, Living Reviews in Solar Physics, 16,
  4

\bibitem[{{Gilliland} {et~al.}(2010){Gilliland}, {Jenkins}, {Borucki},
  {Bryson}, {Caldwell}, {Clarke}, {Dotson}, {Haas}, {Hall}, {Klaus}, {Koch},
  {McCauliff}, {Quintana}, {Twicken}, \& {van Cleve}}]{Gilliland2010}
{Gilliland}, R.~L., {Jenkins}, J.~M., {Borucki}, W.~J., {et~al.} 2010, \apjl,
  713, L160

\bibitem[{Handberg \& Lund(2014)}]{Handberg2014}
Handberg, R. \& Lund, M.~N. 2014, Monthly Notices of the Royal Astronomical
  Society, 445, 2698

\bibitem[{Handberg \& Lund(2019)}]{Handberg2019}
Handberg, R. \& Lund, M.~N. 2019

\bibitem[{{Howell} {et~al.}(2014){Howell}, {Sobeck}, {Haas}, {Still},
  {Barclay}, {Mullally}, {Troeltzsch}, {Aigrain}, {Bryson}, {Caldwell},
  {Chaplin}, {Cochran}, {Huber}, {Marcy}, {Miglio}, {Najita}, {Smith},
  {Twicken}, \& {Fortney}}]{Howell2014}
{Howell}, S.~B., {Sobeck}, C., {Haas}, M., {et~al.} 2014, \pasp, 126, 398

\bibitem[{{Huber} {et~al.}(2011){Huber}, {Bedding}, {Stello}, {Hekker},
  {Mathur}, {Mosser}, {Verner}, {Bonanno}, {Buzasi}, {Campante}, {Elsworth},
  {Hale}, {Kallinger}, {Silva Aguirre}, {Chaplin}, {De Ridder}, {Garc{\'\i}a},
  {Appourchaux}, {Frandsen}, {Houdek}, {Molenda-{\.Z}akowicz}, {Monteiro},
  {Christensen-Dalsgaard}, {Gilliland}, {Kawaler}, {Kjeldsen}, {Broomhall},
  {Corsaro}, {Salabert}, {Sanderfer}, {Seader}, \& {Smith}}]{Huber2011}
{Huber}, D., {Bedding}, T.~R., {Stello}, D., {et~al.} 2011, \apj, 743, 143

\bibitem[{{Huber} {et~al.}(2016){Huber}, {Bryson}, {Haas}, {Barclay},
  {Barentsen}, {Howell}, {Sharma}, {Stello}, \& {Thompson}}]{Huber2016}
{Huber}, D., {Bryson}, S.~T., {Haas}, M.~R., {et~al.} 2016, \apjs, 224, 2

\bibitem[{{Jenkins} {et~al.}(2010){Jenkins}, {Caldwell}, {Chandrasekaran},
  {Twicken}, {Bryson}, {Quintana}, {Clarke}, {Li}, {Allen}, {Tenenbaum}, {Wu},
  {Klaus}, {Van Cleve}, {Dotson}, {Haas}, {Gilliland}, {Koch}, \&
  {Borucki}}]{Jenkins2010}
{Jenkins}, J.~M., {Caldwell}, D.~A., {Chandrasekaran}, H., {et~al.} 2010,
  \apjl, 713, L120

\bibitem[{{Kallinger} {et~al.}(2014){Kallinger}, {De Ridder}, {Hekker},
  {Mathur}, {Mosser}, {Gruberbauer}, {Garc{\'\i}a}, {Karoff}, \&
  {Ballot}}]{Kallinger2014}
{Kallinger}, T., {De Ridder}, J., {Hekker}, S., {et~al.} 2014, \aap, 570, A41

\bibitem[{{Kallinger} {et~al.}(2016){Kallinger}, {Hekker}, {Garcia}, {Huber},
  \& {Matthews}}]{Kallinger2016}
{Kallinger}, T., {Hekker}, S., {Garcia}, R.~A., {Huber}, D., \& {Matthews},
  J.~M. 2016, Science Advances, 2, 1500654

\bibitem[{{Kallinger} {et~al.}(2010){Kallinger}, {Mosser}, {Hekker}, {Huber},
  {Stello}, {Mathur}, {Basu}, {Bedding}, {Chaplin}, {De Ridder}, {Elsworth},
  {Frandsen}, {Garc{\'\i}a}, {Gruberbauer}, {Matthews}, {Borucki}, {Bruntt},
  {Christensen-Dalsgaard}, {Gilliland}, {Kjeldsen}, \& {Koch}}]{Kallinger2010}
{Kallinger}, T., {Mosser}, B., {Hekker}, S., {et~al.} 2010, \aap, 522, A1

\bibitem[{{Karoff}(2012)}]{Karoff2012}
{Karoff}, C. 2012, \mnras, 421, 3170

\bibitem[{{Karoff} {et~al.}(2013){Karoff}, {Campante}, {Ballot}, {Kallinger},
  {Gruberbauer}, {Garc{\'\i}a}, {Caldwell}, {Christiansen}, \&
  {Kinemuchi}}]{Karoff2013}
{Karoff}, C., {Campante}, T.~L., {Ballot}, J., {et~al.} 2013, \apj, 767, 34

\bibitem[{{Kjeldsen} \& {Bedding}(1995)}]{KjeldsenBedding1995}
{Kjeldsen}, H. \& {Bedding}, T.~R. 1995, \aap, 293, 87

\bibitem[{{Kjeldsen} {et~al.}(2010){Kjeldsen}, {Christensen-Dalsgaard},
  {Handberg}, {Brown}, {Gilliland}, {Borucki}, \& {Koch}}]{Kjeldsen2010}
{Kjeldsen}, H., {Christensen-Dalsgaard}, J., {Handberg}, R., {et~al.} 2010,
  Astronomische Nachrichten, 331, 966

\bibitem[{Kobulnicky {et~al.}(2003)Kobulnicky, Nordsieck, Burgh, Smith,
  Percival, Williams, \& O'Donoghue}]{Kobulnicky2003}
Kobulnicky, H.~A., Nordsieck, K.~H., Burgh, E.~B., {et~al.} 2003, in Instrument
  Design and Performance for Optical/Infrared Ground-based Telescopes, ed.
  M.~Iye \& A.~F.~M. Moorwood, Vol. 4841, International Society for Optics and
  Photonics (SPIE), 1634 -- 1644

\bibitem[{{Koch} {et~al.}(2010){Koch}, {Borucki}, {Basri}, {Batalha}, {Brown},
  {Caldwell}, {Christensen-Dalsgaard}, {Cochran}, {DeVore}, {Dunham},
  {Gautier}, {Geary}, {Gilliland}, {Gould}, {Jenkins}, {Kondo}, {Latham},
  {Lissauer}, {Marcy}, {Monet}, {Sasselov}, {Boss}, {Brownlee}, {Caldwell},
  {Dupree}, {Howell}, {Kjeldsen}, {Meibom}, {Morrison}, {Owen}, {Reitsema},
  {Tarter}, {Bryson}, {Dotson}, {Gazis}, {Haas}, {Kolodziejczak}, {Rowe}, {Van
  Cleve}, {Allen}, {Chand rasekaran}, {Clarke}, {Li}, {Quintana}, {Tenenbaum},
  {Twicken}, \& {Wu}}]{Koch2010}
{Koch}, D.~G., {Borucki}, W.~J., {Basri}, G., {et~al.} 2010, \apjl, 713, L79

\bibitem[{{Lomb}(1976)}]{Lomb1976}
{Lomb}, N.~R. 1976, \apss, 39, 447

\bibitem[{{Luger} {et~al.}(2016){Luger}, {Agol}, {Kruse}, {Barnes}, {Becker},
  {Foreman-Mackey}, \& {Deming}}]{Luger2016}
{Luger}, R., {Agol}, E., {Kruse}, E., {et~al.} 2016, \aj, 152, 100

\bibitem[{{Luger} {et~al.}(2017){Luger}, {Foreman-Mackey}, \&
  {Hogg}}]{Luger2017}
{Luger}, R., {Foreman-Mackey}, D., \& {Hogg}, D.~W. 2017, Research Notes of the
  American Astronomical Society, 1, 7

\bibitem[{{Lund} {et~al.}(2017){Lund}, {Silva Aguirre}, {Davies}, {Chaplin},
  {Christensen-Dalsgaard}, {Houdek}, {White}, {Bedding}, {Ball}, {Huber},
  {Antia}, {Lebreton}, {Latham}, {Handberg}, {Verma}, {Basu}, {Casagrande},
  {Justesen}, {Kjeldsen}, \& {Mosumgaard}}]{Lund2017}
{Lund}, M.~N., {Silva Aguirre}, V., {Davies}, G.~R., {et~al.} 2017, \apj, 835,
  172

\bibitem[{{Mathur} {et~al.}(2011{\natexlab{a}}){Mathur}, {Handberg},
  {Campante}, {Garc{\'\i}a}, {Appourchaux}, {Bedding}, {Mosser}, {Chaplin},
  {Ballot}, {Benomar}, {Bonanno}, {Corsaro}, {Gaulme}, {Hekker}, {R{\'e}gulo},
  {Salabert}, {Verner}, {White}, {Brand{\~a}o}, {Creevey}, {Do{\v{g}}an},
  {Elsworth}, {Huber}, {Hale}, {Houdek}, {Karoff}, {Metcalfe},
  {Molenda-{\.Z}akowicz}, {Monteiro}, {Thompson}, {Christensen-Dalsgaard},
  {Gilliland }, {Kawaler}, {Kjeldsen}, {Quintana}, {Sanderfer}, \&
  {Seader}}]{Mathur2011a}
{Mathur}, S., {Handberg}, R., {Campante}, T.~L., {et~al.} 2011{\natexlab{a}},
  \apj, 733, 95

\bibitem[{{Mathur} {et~al.}(2011{\natexlab{b}}){Mathur}, {Hekker},
  {Trampedach}, {Ballot}, {Kallinger}, {Buzasi}, {Garc{\'\i}a}, {Huber},
  {Jim{\'e}nez}, {Mosser}, {Bedding}, {Elsworth}, {R{\'e}gulo}, {Stello},
  {Chaplin}, {De Ridder}, {Hale}, {Kinemuchi}, {Kjeldsen}, {Mullally}, \&
  {Thompson}}]{Mathur2011b}
{Mathur}, S., {Hekker}, S., {Trampedach}, R., {et~al.} 2011{\natexlab{b}},
  \apj, 741, 119

\bibitem[{{Michel} {et~al.}(2008){Michel}, {Baglin}, {Auvergne}, {Catala},
  {Samadi}, {Baudin}, {Appourchaux}, {Barban}, {Weiss}, {Berthomieu},
  {Boumier}, {Dupret}, {Garcia}, {Fridlund}, {Garrido}, {Goupil}, {Kjeldsen},
  {Lebreton}, {Mosser}, {Grotsch-Noels}, {Janot-Pacheco}, {Provost},
  {Roxburgh}, {Thoul}, {Toutain}, {Tiph{\`e}ne}, {Turck-Chieze}, {Vauclair},
  {Vauclair}, {Aerts}, {Alecian}, {Ballot}, {Charpinet}, {Hubert},
  {Ligni{\`e}res}, {Mathias}, {Monteiro}, {Neiner}, {Poretti}, {Renan de
  Medeiros}, {Ribas}, {Rieutord}, {Cort{\'e}s}, \& {Zwintz}}]{michel2008}
{Michel}, E., {Baglin}, A., {Auvergne}, M., {et~al.} 2008, Science, 322, 558

\bibitem[{{Mosser} {et~al.}(2012){Mosser}, {Elsworth}, {Hekker}, {Huber},
  {Kallinger}, {Mathur}, {Belkacem}, {Goupil}, {Samadi}, {Barban}, {Bedding},
  {Chaplin}, {Garc{\'\i}a}, {Stello}, {De Ridder}, {Middour}, {Morris}, \&
  {Quintana}}]{Mosser2012}
{Mosser}, B., {Elsworth}, Y., {Hekker}, S., {et~al.} 2012, \aap, 537, A30

\bibitem[{Oliphant(2006)}]{numpy}
Oliphant, T.~E. 2006, A guide to NumPy, Vol.~1 (Trelgol Publishing USA)

\bibitem[{{Palla} \& {Stahler}(1990)}]{Palla1990}
{Palla}, F. \& {Stahler}, S.~W. 1990, \apjl, 360, L47

\bibitem[{{Pande} {et~al.}(2018){Pande}, {Bedding}, {Huber}, \&
  {Kjeldsen}}]{Pande2018}
{Pande}, D., {Bedding}, T.~R., {Huber}, D., \& {Kjeldsen}, H. 2018, \mnras,
  480, 467

\bibitem[{{Paxton} {et~al.}(2011){Paxton}, {Bildsten}, {Dotter}, {Herwig},
  {Lesaffre}, \& {Timmes}}]{Paxton2011}
{Paxton}, B., {Bildsten}, L., {Dotter}, A., {et~al.} 2011, \apjs, 192, 3

\bibitem[{{Paxton} {et~al.}(2013){Paxton}, {Cantiello}, {Arras}, {Bildsten},
  {Brown}, {Dotter}, {Mankovich}, {Montgomery}, {Stello}, {Timmes}, \&
  {Townsend}}]{Paxton2013}
{Paxton}, B., {Cantiello}, M., {Arras}, P., {et~al.} 2013, \apjs, 208, 4

\bibitem[{{Paxton} {et~al.}(2015){Paxton}, {Marchant}, {Schwab}, {Bauer},
  {Bildsten}, {Cantiello}, {Dessart}, {Farmer}, {Hu}, {Langer}, {Townsend},
  {Townsley}, \& {Timmes}}]{Paxton2015}
{Paxton}, B., {Marchant}, P., {Schwab}, J., {et~al.} 2015, \apjs, 220, 15

\bibitem[{{Paxton} {et~al.}(2018){Paxton}, {Schwab}, {Bauer}, {Bildsten},
  {Blinnikov}, {Duffell}, {Farmer}, {Goldberg}, {Marchant}, {Sorokina},
  {Thoul}, {Townsend}, \& {Timmes}}]{Paxton2018}
{Paxton}, B., {Schwab}, J., {Bauer}, E.~B., {et~al.} 2018, \apjs, 234, 34

\bibitem[{{Paxton} {et~al.}(2019){Paxton}, {Smolec}, {Schwab}, {Gautschy},
  {Bildsten}, {Cantiello}, {Dotter}, {Farmer}, {Goldberg}, {Jermyn}, {Kanbur},
  {Marchant}, {Thoul}, {Townsend}, {Wolf}, {Zhang}, \& {Timmes}}]{Paxton2019}
{Paxton}, B., {Smolec}, R., {Schwab}, J., {et~al.} 2019, \apjs, 243, 10

\bibitem[{{Pecaut} \& {Mamajek}(2016)}]{Pecaut2016}
{Pecaut}, M.~J. \& {Mamajek}, E.~E. 2016, \mnras, 461, 794

\bibitem[{{Pinheiro}(2008)}]{Pinheiro2008}
{Pinheiro}, F.~J.~G. 2008, \aap, 478, 193

\bibitem[{{Pinsonneault} {et~al.}(2014){Pinsonneault}, {Elsworth}, {Epstein},
  {Hekker}, {M{\'e}sz{\'a}ros}, {Chaplin}, {Johnson}, {Garc{\'\i}a},
  {Holtzman}, {Mathur}, {Garc{\'\i}a P{\'e}rez}, {Silva Aguirre}, {Girardi},
  {Basu}, {Shetrone}, {Stello}, {Allende Prieto}, {An}, {Beck}, {Beers},
  {Bizyaev}, {Bloemen}, {Bovy}, {Cunha}, {De Ridder}, {Frinchaboy},
  {Garc{\'\i}a-Hern{\'a}ndez}, {Gilliland}, {Harding}, {Hearty}, {Huber},
  {Ivans}, {Kallinger}, {Majewski}, {Metcalfe}, {Miglio}, {Mosser}, {Muna},
  {Nidever}, {Schneider}, {Serenelli}, {Smith}, {Tayar}, {Zamora}, \&
  {Zasowski}}]{Pinsonneault2014}
{Pinsonneault}, M.~H., {Elsworth}, Y., {Epstein}, C., {et~al.} 2014, \apjs,
  215, 19

\bibitem[{{Piskunov} \& {Valenti}(2017)}]{Piskunov2017}
{Piskunov}, N. \& {Valenti}, J.~A. 2017, \aap, 597, A16

\bibitem[{{Preibisch} {et~al.}(2001){Preibisch}, {Guenther}, \&
  {Zinnecker}}]{Preibisch2001}
{Preibisch}, T., {Guenther}, E., \& {Zinnecker}, H. 2001, \aj, 121, 1040

\bibitem[{{Price-Whelan} {et~al.}(2018){Price-Whelan}, {Sip{\H{o}}cz},
  {G{\"u}nther}, {Lim}, {Crawford}, {Conseil}, {Shupe}, {Craig}, {Dencheva},
  {Ginsburg}, {VanderPlas}, {Bradley}, {P{\'e}rez-Su{\'a}rez}, {de Val-Borro},
  {Paper Contributors}, {Aldcroft}, {Cruz}, {Robitaille}, {Tollerud},
  {Coordination Committee}, {Ardelean}, {Babej}, {Bach}, {Bachetti}, {Bakanov},
  {Bamford}, {Barentsen}, {Barmby}, {Baumbach}, {Berry}, {Biscani}, {Boquien},
  {Bostroem}, {Bouma}, {Brammer}, {Bray}, {Breytenbach}, {Buddelmeijer},
  {Burke}, {Calderone}, {Cano Rodr{\'\i}guez}, {Cara}, {Cardoso}, {Cheedella},
  {Copin}, {Corrales}, {Crichton}, {D{\textquoteright}Avella}, {Deil},
  {Depagne}, {Dietrich}, {Donath}, {Droettboom}, {Earl}, {Erben}, {Fabbro},
  {Ferreira}, {Finethy}, {Fox}, {Garrison}, {Gibbons}, {Goldstein}, {Gommers},
  {Greco}, {Greenfield}, {Groener}, {Grollier}, {Hagen}, {Hirst}, {Homeier},
  {Horton}, {Hosseinzadeh}, {Hu}, {Hunkeler}, {Ivezi{\'c}}, {Jain}, {Jenness},
  {Kanarek}, {Kendrew}, {Kern}, {Kerzendorf}, {Khvalko}, {King}, {Kirkby},
  {Kulkarni}, {Kumar}, {Lee}, {Lenz}, {Littlefair}, {Ma}, {Macleod},
  {Mastropietro}, {McCully}, {Montagnac}, {Morris}, {Mueller}, {Mumford},
  {Muna}, {Murphy}, {Nelson}, {Nguyen}, {Ninan}, {N{\"o}the}, {Ogaz}, {Oh},
  {Parejko}, {Parley}, {Pascual}, {Patil}, {Patil}, {Plunkett}, {Prochaska},
  {Rastogi}, {Reddy Janga}, {Sabater}, {Sakurikar}, {Seifert}, {Sherbert},
  {Sherwood-Taylor}, {Shih}, {Sick}, {Silbiger}, {Singanamalla}, {Singer},
  {Sladen}, {Sooley}, {Sornarajah}, {Streicher}, {Teuben}, {Thomas},
  {Tremblay}, {Turner}, {Terr{\'o}n}, {van Kerkwijk}, {de la Vega}, {Watkins},
  {Weaver}, {Whitmore}, {Woillez}, {Zabalza}, \& {Contributors}}]{astropy:2018}
{Price-Whelan}, A.~M., {Sip{\H{o}}cz}, B.~M., {G{\"u}nther}, H.~M., {et~al.}
  2018, \aj, 156, 123

\bibitem[{{Rauer} {et~al.}(2014){Rauer}, {Catala}, {Aerts}, {Appourchaux},
  {Benz}, {Brandeker}, {Christensen-Dalsgaard}, {Deleuil}, {Gizon}, {Goupil},
  {G{\"u}del}, {Janot-Pacheco}, {Mas-Hesse}, {Pagano}, {Piotto}, {Pollacco},
  {Santos}, {Smith}, {Su{\'a}rez}, {Szab{\'o}}, {Udry}, {Adibekyan}, {Alibert},
  {Almenara}, {Amaro-Seoane}, {Eiff}, {Asplund}, {Antonello}, {Barnes},
  {Baudin}, {Belkacem}, {Bergemann}, {Bihain}, {Birch}, {Bonfils}, {Boisse},
  {Bonomo}, {Borsa}, {Brand {\~a}o}, {Brocato}, {Brun}, {Burleigh}, {Burston},
  {Cabrera}, {Cassisi}, {Chaplin}, {Charpinet}, {Chiappini}, {Church},
  {Csizmadia}, {Cunha}, {Damasso}, {Davies}, {Deeg}, {D{\'\i}az}, {Dreizler},
  {Dreyer}, {Eggenberger}, {Ehrenreich}, {Eigm{\"u}ller}, {Erikson}, {Farmer},
  {Feltzing}, {de Oliveira Fialho}, {Figueira}, {Forveille}, {Fridlund},
  {Garc{\'\i}a}, {Giommi}, {Giuffrida}, {Godolt}, {Gomes da Silva}, {Granzer},
  {Grenfell}, {Grotsch-Noels}, {G{\"u}nther}, {Haswell}, {Hatzes},
  {H{\'e}brard}, {Hekker}, {Helled}, {Heng}, {Jenkins}, {Johansen},
  {Khodachenko}, {Kislyakova}, {Kley}, {Kolb}, {Krivova}, {Kupka}, {Lammer},
  {Lanza}, {Lebreton}, {Magrin}, {Marcos-Arenal}, {Marrese}, {Marques},
  {Martins}, {Mathis}, {Mathur}, {Messina}, {Miglio}, {Montalban}, {Montalto},
  {Monteiro}, {Moradi}, {Moravveji}, {Mordasini}, {Morel}, {Mortier},
  {Nascimbeni}, {Nelson}, {Nielsen}, {Noack}, {Norton}, {Ofir}, {Oshagh},
  {Ouazzani}, {P{\'a}pics}, {Parro}, {Petit}, {Plez}, {Poretti}, {Quirrenbach},
  {Ragazzoni}, {Raimondo}, {Rainer}, {Reese}, {Redmer}, {Reffert},
  {Rojas-Ayala}, {Roxburgh}, {Salmon}, {Santerne}, {Schneider}, {Schou},
  {Schuh}, {Schunker}, {Silva-Valio}, {Silvotti}, {Skillen}, {Snellen}, {Sohl},
  {Sousa}, {Sozzetti}, {Stello}, {Strassmeier}, {{\v{S}}vanda}, {Szab{\'o}},
  {Tkachenko}, {Valencia}, {Van Grootel}, {Vauclair}, {Ventura}, {Wagner},
  {Walton}, {Weingrill}, {Werner}, {Wheatley}, \& {Zwintz}}]{Rauer2014}
{Rauer}, H., {Catala}, C., {Aerts}, C., {et~al.} 2014, Experimental Astronomy,
  38, 249

\bibitem[{{Ricker} {et~al.}(2014){Ricker}, {Vanderspek}, {Latham}, \&
  {Winn}}]{Ricker2014}
{Ricker}, G.~R., {Vanderspek}, R.~K., {Latham}, D.~W., \& {Winn}, J.~N. 2014,
  in American Astronomical Society Meeting Abstracts, Vol. 224, American
  Astronomical Society Meeting Abstracts \#224, 113.02

\bibitem[{{Samadi} {et~al.}(2005){Samadi}, {Goupil}, {Alecian}, {Baudin},
  {Georgobiani}, {Trampedach}, {Stein}, \& {Nordlund}}]{Samadi2005}
{Samadi}, R., {Goupil}, M.~J., {Alecian}, E., {et~al.} 2005, Journal of
  Astrophysics and Astronomy, 26, 171

\bibitem[{{S{\'a}nchez-Bl{\'a}zquez} {et~al.}(2006){S{\'a}nchez-Bl{\'a}zquez},
  {Peletier}, {Jim{\'e}nez-Vicente}, {Cardiel}, {Cenarro},
  {Falc{\'o}n-Barroso}, {Gorgas}, {Selam}, \& {Vazdekis}}]{Sanchez2006}
{S{\'a}nchez-Bl{\'a}zquez}, P., {Peletier}, R.~F., {Jim{\'e}nez-Vicente}, J.,
  {et~al.} 2006, \mnras, 371, 703

\bibitem[{{Savitzky} \& {Golay}(1964)}]{Savitzky1964}
{Savitzky}, A. \& {Golay}, M.~J.~E. 1964, Analytical Chemistry, 36, 1627

\bibitem[{{Scargle}(1982)}]{Scargle1982}
{Scargle}, J.~D. 1982, \apj, 263, 835

\bibitem[{{Schofield} {et~al.}(2019){Schofield}, {Chaplin}, {Huber},
  {Campante}, {Davies}, {Miglio}, {Ball}, {Appourchaux}, {Basu}, {Bedding},
  {Christensen-Dalsgaard}, {Creevey}, {Garc{\'\i}a}, {Handberg}, {Kawaler},
  {Kjeldsen}, {Latham}, {Lund}, {Metcalfe}, {Ricker}, {Serenelli}, {Silva
  Aguirre}, {Stello}, \& {Vanderspek}}]{Schofield2019}
{Schofield}, M., {Chaplin}, W.~J., {Huber}, D., {et~al.} 2019, \apjs, 241, 12

\bibitem[{{Shaw} {et~al.}(2007){Shaw}, {Bridges}, \& {Hobson}}]{Shaw2007}
{Shaw}, J.~R., {Bridges}, M., \& {Hobson}, M.~P. 2007, \mnras, 378, 1365

\bibitem[{{Skilling}(2006)}]{Skilling2006}
{Skilling}, J. 2006, in American Institute of Physics Conference Series, Vol.
  872, Bayesian Inference and Maximum Entropy Methods In Science and
  Engineering, ed. A.~{Mohammad-Djafari}, 321--330

\bibitem[{{Stassun} {et~al.}(2018){Stassun}, {Corsaro}, {Pepper}, \&
  {Gaudi}}]{Stassun2018}
{Stassun}, K.~G., {Corsaro}, E., {Pepper}, J.~A., \& {Gaudi}, B.~S. 2018, \aj,
  155, 22

\bibitem[{{Stello} {et~al.}(2008){Stello}, {Bruntt}, {Preston}, \&
  {Buzasi}}]{Stello2008}
{Stello}, D., {Bruntt}, H., {Preston}, H., \& {Buzasi}, D. 2008, \apjl, 674,
  L53

\bibitem[{{Stello} {et~al.}(2009){Stello}, {Chaplin}, {Basu}, {Elsworth}, \&
  {Bedding}}]{Stello2009}
{Stello}, D., {Chaplin}, W.~J., {Basu}, S., {Elsworth}, Y., \& {Bedding}, T.~R.
  2009, \mnras, 400, L80

\bibitem[{{Trotta}(2006)}]{Trotta2006}
{Trotta}, R. 2006, in Statistical Problems in Particle Physics, Astrophysics
  and Cosmology, ed. L.~{Lyons} \& M.~{Karag{\"o}z {\"U}nel}, 15

\bibitem[{{Valenti} \& {Piskunov}(1996)}]{Valenti1996}
{Valenti}, J.~A. \& {Piskunov}, N. 1996, \aaps, 118, 595

\bibitem[{{Valenti} {et~al.}(1998){Valenti}, {Piskunov}, \&
  {Johns-Krull}}]{Valenti1998}
{Valenti}, J.~A., {Piskunov}, N., \& {Johns-Krull}, C.~M. 1998, \apj, 498, 851

\bibitem[{{Virtanen} {et~al.}(2020){Virtanen}, {Gommers}, {Oliphant},
  {Haberland}, {Reddy}, {Cournapeau}, {Burovski}, {Peterson}, {Weckesser},
  {Bright}, {van der Walt}, {Brett}, {Wilson}, {Jarrod Millman}, {Mayorov},
  {Nelson}, {Jones}, {Kern}, {Larson}, {Carey}, {Polat}, {Feng}, {Moore}, {Vand
  erPlas}, {Laxalde}, {Perktold}, {Cimrman}, {Henriksen}, {Quintero}, {Harris},
  {Archibald}, {Ribeiro}, {Pedregosa}, {van Mulbregt}, \&
  {Contributors}}]{scipy}
{Virtanen}, P., {Gommers}, R., {Oliphant}, T.~E., {et~al.} 2020, Nature Methods

\bibitem[{{Walker} {et~al.}(2003){Walker}, {Matthews}, {Kuschnig}, {Johnson},
  {Rucinski}, {Pazder}, {Burley}, {Walker}, {Skaret}, {Zee}, {Grocott},
  {Carroll}, {Sinclair}, {Sturgeon}, \& {Harron}}]{walker03}
{Walker}, G., {Matthews}, J., {Kuschnig}, R., {et~al.} 2003, \pasp, 115, 1023

\bibitem[{{White} \& {Basri}(2003)}]{White2003}
{White}, R. \& {Basri}, G. 2003, in IAU Symposium, Vol. 211, Brown Dwarfs, ed.
  E.~{Mart{\'\i}n}, 143

\end{thebibliography}

\end{document}